\documentclass[useAMS,usenatbib]{mn2e}
\usepackage{graphicx}
\usepackage[abs]{overpic}
\usepackage[fleqn]{amsmath}
\usepackage{color}
\usepackage{xspace}
\newcommand{\note}[1]{}
\newcommand{\snotejan}[1]{}
\newcommand{\knotejan}[1]{}
\newcommand{\ksec}[1]{}
\newcommand{\rp}{r_\rmn{p}}

\newcommand{\Nc}{N_\rmn{c}}
\newcommand{\fpair}{f_\rmn{p}}
\newcommand{\fcontrol}{f_\rmn{c}}
\newcommand{\hkpc}{$h^{-1}_{70}$ kpc\xspace}
\newcommand{\kms}{km s$^{-1}$\xspace}

\title[Galaxy Zoo: Quantifying Morphological Indicators of Galaxy Interaction]{Galaxy Zoo: Quantifying Morphological Indicators of Galaxy Interaction\thanks{This publication has been made possible by the participation of more than 200,000 volunteers in the Galaxy Zoo project. Their contributions are acknowledged at http://authors.galaxyzoo.org}}
\author[K. R. V. Casteels et al.]{Kevin R. V. Casteels,$^{1,2}$\thanks{E-mail: kcasteels@am.ub.es (KRVC)}
Steven P. Bamford,$^{2}$\thanks{E-mail: steven.bamford@nottingham.ac.uk (SPB)}
Ramin A. Skibba,$^{3}$\newauthor
Karen L. Masters,$^{4,5}$
Chris J. Lintott,$^{6,7}$
William C. Keel,$^{8}$
Kevin Schawinski,$^{9,10,11}$\newauthor
Robert C. Nichol,$^{4,5}$
Arfon M. Smith$^{7}$
\smallskip\\
$^{1}$Departament d'Astronomia i Meteorologia, Universitat de Barcelona, Mart\'{i} Franques 1, E-08028 Barcelona, Espa\~{n}a\\
$^{2}$School of Physics and Astronomy, The University of Nottingham, University Park, Nottingham, NG7 2RD, UK\\
$^{3}$Steward Observatory, University of Arizona, 933 N. Cherry Ave., Tucson, AZ 85721, USA\\
$^{4}$Institute of Cosmology \& Gravitation, University of Portsmouth, Dennis Sciama Building, Portsmouth, PO1 3FX, UK\\
$^{5}$South East Physics Network, http://www.sepnet.ac.uk\\
$^{6}$Oxford Astrophysics, Department of Physics, University of Oxford, Denys Wilkinson Building, Keble Road, Oxford, OX1 3RH, UK\\
$^{7}$Astronomy Department, Adler Planetarium and Astronomy Museum, 1300 Lake Shore Drive, Chicago, IL 60605, USA\\
$^{8}$Department of Physics \& Astronomy, University of Alabama, 206 Gallalee Hall, 514 University Boulevard, Tuscaloosa, AL 35487-0234, USA\\
$^{9}$Department of Physics, Yale University, New Haven, CT 06511, USA\\
$^{10}$Yale Center for Astronomy and Astrophysics, Yale University, P.O. Box 208121, New Haven, CT 06520, USA\\
$^{11}$Einstein Fellow}
\begin{document}

\date{Accepted 12 November 2012. Received 29 October 2012; in original form 21 June 2012}

\pagerange{\pageref{firstpage}--\pageref{lastpage}} \pubyear{2012}

\maketitle

\label{firstpage}

\begin{abstract}

We use Galaxy Zoo 2 visual classifications to study the morphological signatures of interaction between similar-mass galaxy pairs in the Sloan Digital Sky Survey.  We find that many observable features correlate with projected pair separation; not only obvious indicators of merging, disturbance and tidal tails, but also more regular features, such as spiral arms and bars.  These trends are robustly quantified, using a control sample to account for observational biases, producing measurements of the strength and separation scale of various morphological responses to pair interaction.  For example, we find that the presence of spiral features is enhanced at scales $\la 70$ \hkpc, probably due to both increased star formation and the formation of tidal tails.  On the other hand, the likelihood of identifying a bar decreases significantly in pairs with separations $\la 30$ \hkpc, suggesting that bars are suppressed by close interactions between galaxies of similar mass.

We go on to show how morphological indicators of physical interactions provide a way of significantly refining standard estimates for the frequency of close pair interactions, based on velocity offset and projected separation.  The presence of loosely wound spiral arms is found to be a particularly reliable signal of an interaction, for projected pair separations up to $\sim 100$ \hkpc.  We use this indicator to demonstrate our method, constraining the fraction of low-redshift galaxies in truly interacting pairs, with $M_* > 10^{9.5} M_{\sun}$ and mass ratio $< 4$, to be between 0.4 -- 2.7 per cent.

\end{abstract}


\section{Introduction}\label{intro}

When galaxies approach one another, their mutual gravitational attraction can result in substantial disruptions to their morphologies, such as tidal arms, counter arms, bridges and tails. 
Many studies have shown, both analytically and using numerical simulations, that galaxies of 
similar mass can provoke dramatic disturbances in the stellar distributions of one another, with the details depending on the orbital parameters of the interaction (e.g., \citealt{toom1972,barn1992,howa1993,gerb1994,barn2011}).  Gravitational perturbations can also redistribute the gas content of galaxies, potentially leading to changes in their star-formation properties (e.g., \citealt{nogu1988,barn1996}).  These effects are strong functions of pair separation, and hence should be most obvious after the galaxies' first pass, and particularly around times of closest approach.  Many interactions are ultimately likely to result in the complete morphological transformation of the galaxies involved (e.g., \citealt{toom1977,hopk2008}).  However, more subtle effects are expected both earlier, while the galaxies are on their initial approach, and at times of wide separation between passes (e.g., \citealt{pere2006,lotz2008,stru2011}).

The expected strength and prevalence of pair interactions mean they are potentially important for determining the properties and evolution of the galaxy population.  It is therefore critical that we test our theoretical expectations of the effects of such interactions by studying representative samples of interacting systems.  Furthermore, we may utilise the observed frequency of interacting pairs to discriminate between the details of cosmological galaxy formation models.

A relatively straightforward, and physically motivated, definition of `interacting' galaxies is a pair for which the tidal force experienced across one of the galaxies, averaged over its internal dynamical time, $F_\rmn{t}$, is at least some specified fraction of the gravitational force binding the outer regions of that galaxy, $F_\rmn{g}$.  By this definition, all interacting pairs should produce significant internal dynamical effects, which would have otherwise been absent, in at least one of the member galaxies.  In practice, however, it is difficult to measure the forces involved.  We may estimate them by studying the effects of an interaction, but the relative orientations and types of the galaxies in each pair, as well as observational limitations, lead to large variations in the apparent effects for interactions of a given strength, $F_\rmn{t}/F_\rmn{g}$.

A more convenient definition, which is roughly equivalent, although only statistically applicable, is that 
%a pair are `interacting' if their gravitational influence upon one another would have observable effects, if only their orientations and constituent galaxy types were favourable for producing such effects.  
a pair are `interacting' if their gravitational influence upon one another could have observable effects in a favourable orientation and mix of galaxy types.
For example, a pair of elliptical galaxies might not display signatures of an interaction in a given observation, but would still count as `interacting' by this definition if the tidal forces they are experiencing would have been sufficient to produce an observable signature in a pair of spirals.  The observational details of a particular dataset therefore fix the minimum $F_\rmn{t}/F_\rmn{g}$ probed.  This definition removes much of the incompleteness associated with only considering pairs with observational signs of interaction, but of course only a fraction of such 'interacting' galaxies will possess observational signatures.  As we shall see later in this paper, it is nevertheless possible to constrain the fraction of galaxies interacting according to this definition.

%The above definitions of `interacting' galaxies are somewhat similar to existing attempts to quantify the strength of galaxy interactions, for example the $Q_E$ parameter of \citet{elme1991}, which is based on $Q_D$ parameter of \citet{daha1984} (also see section 7.1 of \citealt{bere2004}).

Studies of galaxy pairs typically discuss close pairs, bound pairs, merging pairs, or pairs with observational disturbances, but often mix the usage and definitions of these classes.  Interacting pairs are closely related to bound pairs, for which the sum of the gravitational potential energy and kinetic energy of the pair is negative.  However, not all interacting pairs are bound, particularly where both are part of a larger system, such as a galaxy cluster.  Likewise, not all bound pairs will be experiencing significant tidal interactions.  Interacting pairs are also closely related to mergers.  Galaxies in bound pairs will typically merge on relatively short time-scales if they are experiencing significant tidal interactions, as the kinetic energy of the pair orbit is transferred to deforming and heating the internal mass distribution of each galaxy (e.g. \citealt{stru1999}).

We can identify close pairs of galaxies which are likely to be sufficiently near to one another such that they are interacting (and potentially bound and will eventually merge) using projected distance and line-of-sight velocity \citep{char1991}.  However, this approach suffers from significant contamination and incompleteness (with respect to the above definitions of physically meaningful interacting, bound or merging selections), due to a lack of full spatial information and the inverse relationship between relative velocity and separation for loose pairs (i.e. very close pairs can have very large relative peculiar velocities, so appear significantly separated in redshift space).  
Observational signatures of interactions, for example visual classifications, quantitative morphological measurements or induced star formation, may be used to improve the selection of truly interacting galaxies.  However, as often one wishes to study the physical effect of interactions, one must be careful to avoid a circular argument.

%***********************
The early atlases of \citet{voro1959,voro1977} and \citet{arp1966} clearly demonstrated that interactions between galaxies can have profound effects on their morphologies, providing examples of bridges, tails, distorted spiral patterns, and other features.  These morphological changes where observed between pairs up to over 100 \hkpc, as is the case for the bridging filament of Arp 295.  The restricted three-body simulations of \citet{toom1972} clearly demonstrated that these features are the result of strong tidal forces between the interacting galaxies.
%********************************
That galaxy interactions can also induce star-formation was first suggested by \citet{lars1978}, who found that the scatter in the UBV colours of interacting galaxies from the Arp atlas was significantly larger than normal galaxies from the \emph{Hubble Atlas} \citep{sand1961}.
Similar evidence for interaction induced star-formation has also been found over a wide range of the energy spectrum, from near-UV to radio (e.g. \citealt{kenn1984,keel1985,bush1986,kenn1987,bush1987,humm1990}). 
%*********************************

Studies which use quantitative measures of morphology find signs of interaction at fairly small projected 
separations. Using the CAS method of \citet{cons2003} to measure galaxy asymmetry ($A$) and concentration ($C$),
\citet{hern2005} found that both these quantities increase, relative to isolated galaxies, for galaxy pairs with separations less than the photometric diameter of the primary. \citet{depr2007} reliably identified interacting pairs with projected separations up to $\rp \la 50$ \hkpc using $A>0.35$ and visual confirmation, for a sample of pairs with line-of-sight velocity differences 
$\Delta V < 500$ \kms.\footnote{Throughout this introduction, separations quoted from other studies have been converted to units of \hkpc as necessary.}  Similarly, \citet{elli2010} show that asymmetry increases for $\rp \la 50$ \hkpc for a sample of pairs with $\Delta V < 200$ \kms.

Meanwhile, studies which probe the effects of tidal interaction through star formation modulations find changes up to larger projected separations.
%The effects of tidal interactions can also be studied by measuring at what separation star formation modulations, typically enhancements, become noticeable.
\citet{niko2004} demonstrate an increase in star formation at at $\rp \la 70$ \hkpc for early and mixed type pairs, and 
at $\rp \la 430$ \hkpc (their maximum separation probed) for late type pairs.
They also find that pairs with $\rp \la 110$ \hkpc show a strong increase in central concentration, suggestive of nuclear starbursts.
%, signifying that the B/T ratio is significantly increasing at this separation.
\citet{li2008} find a star formation increase for close pairs, with star formation rate (SFR) enhanced by a factor of $1.5$ at $\rp \la 140$ \hkpc to a factor of $4$ at $\rp \la 30$ \hkpc. This strong dependence on $\rp$, is contrasted with a weak dependence on luminosity ratio, with the star formation enhancement being strongest in lower luminosity galaxies.  \citet{elli2008}, \citet{roba2009} and \citet{patt2011} all find a strong increase in SFR for $\rp<40$ \hkpc, while \citet{patt2011} also see a smaller increase up to at least $\rp < 80$ \hkpc (their maximum separation probed) for $\Delta V < 200$ \kms pairs. There is also evidence that equivalent levels of tidally induced star formation require smaller $\rp$ in denser environments (e.g., \citealt{lamb2003,alon2004}).

%It is worth noting that similar, though much less extensive results have been found in non-optical wavebands in nearby galaxy samples. This work extends from radio continuum studies dating to the late 1970s to more recent satellite results (e.g., from Spitzer (Smith et al. 2007AJ....133..791) and GALEX (Smith et al. 2010AJ....139.1212, Smith & Struck, 2010AJ....140.1975)). 

Some of the variation in the separation scale at which different studies begin to see the effects of tidal interactions is likely due to differences in the mass and luminosity ranges of the samples, as well as the methods used. However, it appears clear that the effects of tidal interactions are found at larger projected separations when identified by induced star formation (up to $80$ to $100$ \hkpc), compared to quantitative measurements of asymmetry (up to $\sim 30$ \hkpc).  This is consistent with the results of \citet{lotz2008}, which used simulations to show that quantitative morphological methods for finding merging galaxies, such as $A$, Gini and $M_{20}$, are most sensitive for galaxies undergoing close passages and during the post merger phase. Induced star formation, on the other hand, will be evident between passes, when the galaxies achieve a wide separation before falling back towards one another, or in galaxies which have experienced a close encounter but will not merge.  Note that in dry mergers there may be no star-formation signature of the interaction, and morphological features will typically only be observable for short times \citep{bell2006}.

As mentioned previously, interacting galaxies can produce distinctive morphological features such as tidal arms, counter arms, bridges and tails, which are best classified visually.  These features are extremely reliable for discriminating truly interacting galaxies from interlopers in close pair catalogues.  Features such as two loosely-wound tidal arms may not be detectable by quantitative morphology methods because these galaxies may not appear to be sufficiently asymmetric or disturbed, especially between the first and second pass when the galaxies may appear to be widely separated.  One of the advantages of using visual morphological classifications over automated methods is the ability to identify very faint and subtle features. Tidal features are known to become rapidly undetectable as a function of time and survey imaging depth (e.g., \citealt{bell2006,scha2010}), however we find that the Galaxy Zoo classifications used in this paper are extremely sensitive to faint features.  Furthermore, as we will show, by studying the occurrence of such features in a statistical sense, and making weak assumptions concerning the observability of physical interactions, we are able to make decisive statements concerning the prevalence of interactions.

The visual classification of peculiar, disturbed and interacting galaxies has a long history, beginning with \citet{hubb1926}. The catalogues of \citet{voro1959,voro1977}, \citet{sand1961}, and \citet{arp1966} complied together a significant number of galaxies with obvious tidal features.  Such work continues to be valuable today, for example \citet{brid2010} uses visual classifications of galactic bridges and tails in the CFHT Legacy Survey to study the evolution of the galaxy interaction fraction (GIF) with redshift. These galaxies were either isolated merger remnants or fairly close interacting pairs, due to their requirement that galaxy pairs be connected by a bridge.  \citet{naka2003} and \citet{fuku2007} visually classified a subsample of $\sim 2500$ bright galaxies from the Sloan Digital Sky Survey (SDSS; \citealt{york2000}) imaging of SDSS galaxy objects, finding that $\sim 1.5$ per cent of galaxies in their nearby magnitude-limited sample show morphological indications of interaction.  \citet{nair2010} provide an impressive catalogue of detailed visual classifications for over $14\,000$ bright SDSS galaxies, which includes information regarding tidal tails and other indicators of interaction.

The Galaxy Zoo project has enabled visual classification to be performed for extremely large samples, allowing the continued use of this valuable technique with modern surveys.  \citet{skib2009} obtained the marked correlation function for the Galaxy Zoo 1 merger classification likelihood and found that it increases sharply in their closest $\rp$ bin (of 170 \hkpc width), and found evidence that most of this increase was for pairs with $\rp \la 30$ \hkpc.  Taking an alternative approach, \citet{darg2010a} and \citet{darg2010b} imposed thresholds to select Galaxy Zoo 1 galaxies with high merger classification likelihoods and study their frequency and properties.  Most of these galaxies are either highly disturbed systems or very close pairs. While these studies have been successful at identifying a subset of interacting pairs, they primarily select galaxies which have relatively small projected separations and so do not typically identify interacting pairs which are at large projected separation between their first and second close passes.

%(See also Robaina+ 2009 and Robaina & Bell 2010.)

%While many of the pairs selected in this way will be truly interacting, this method suffers from line of sight projections being misinterpreted as mergers.
%We also find, as discussed in section \ref{results}, that galaxies identified as mergers in this way will quite close...
%-To date, we are unaware or any study which explores up to what separation the effects of tidal interactions are visible in galaxy pairs...

%"Studies to select mergers by visual-selection have
%not reached such scales as MOSES until this work.

%Le Fevre et al. (2000) visually examined 285 Hubble images
%and found a merger fraction of 10 ± 2% over 0 < z
%1.2.

In this paper we use visual classifications from Galaxy Zoo 2 to study what morphological changes are taking place in interacting galaxy pairs as a function of physical projected separation ($\rp$) and line of sight velocity difference ($\Delta V$).  We then use these results to estimate the frequency of pair galaxy interactions in the local universe.

In Section \ref{data} we describe the data set and sample selection, in Section \ref{method} we outline the methods employed, in Section \ref{results} we present our results, and in Section \ref{disc} we summarise our results and discuss their implications.  A $\Lambda CDM$ cosmology is assumed throughout, with $\Omega_{\Lambda}$ = 0.7, $\Omega_{m}$ = 0.3, and $h_{70} = H_0 / (70$ \kms Mpc$^{-1})$.

\section{Galaxy Zoo 2 data and sample selection}\label{data}

Although there have been many attempts at completely automating morphological classification, visual inspection remains the preferred method for many astronomers.  However, for the large samples produced by modern surveys, visual classification is not feasible for a normal research team to perform in a reasonable time.  
Galaxy Zoo \citep{lint2008} is an online citizen science project designed to address this problem, by involving large numbers of the public in classifying the morphological features of galaxies. The original Galaxy Zoo website collected classifications for nearly one million galaxies from SDSS Data Release 6 (DR6; \citealt{adel2008}), which were processed to produce catalogues of visual classifications together with estimates of their accuracy.  These data have been released to the public\footnote{Galaxy Zoo data is publicly available at http://data.galaxyzoo.org/.}, and are described in \citet{lint2010}.

The original Galaxy Zoo (GZ1) was limited to coarse morphological classification.  Following its success, a subsequent project was launched, this time collecting much more detailed morphological information via a question tree, for a subset of $\sim300\,000$ of the brightest galaxies from Galaxy Zoo.  This project, named Galaxy Zoo 2 (GZ2), ran from 16 February 2009 until 22 April 2010 and collected $16\,340\,298$ classifications (comprising a total of $58\,719\,719$ questions) by $83\,943$ participants for $325\,651$ galaxy images\footnote{The GZ2 website is archived at http://zoo2.galaxyzoo.org.}.  Both GZ1 and GZ2 used $gri$ composite colour images provided by the SDSS, created following the prescription of \citet{lupton}.  These were displayed such that each galaxy had the same apparent size.  See \citet{mast2011} for additional discussion of the GZ2 dataset.

\begin{figure*}
\includegraphics[width=160mm]{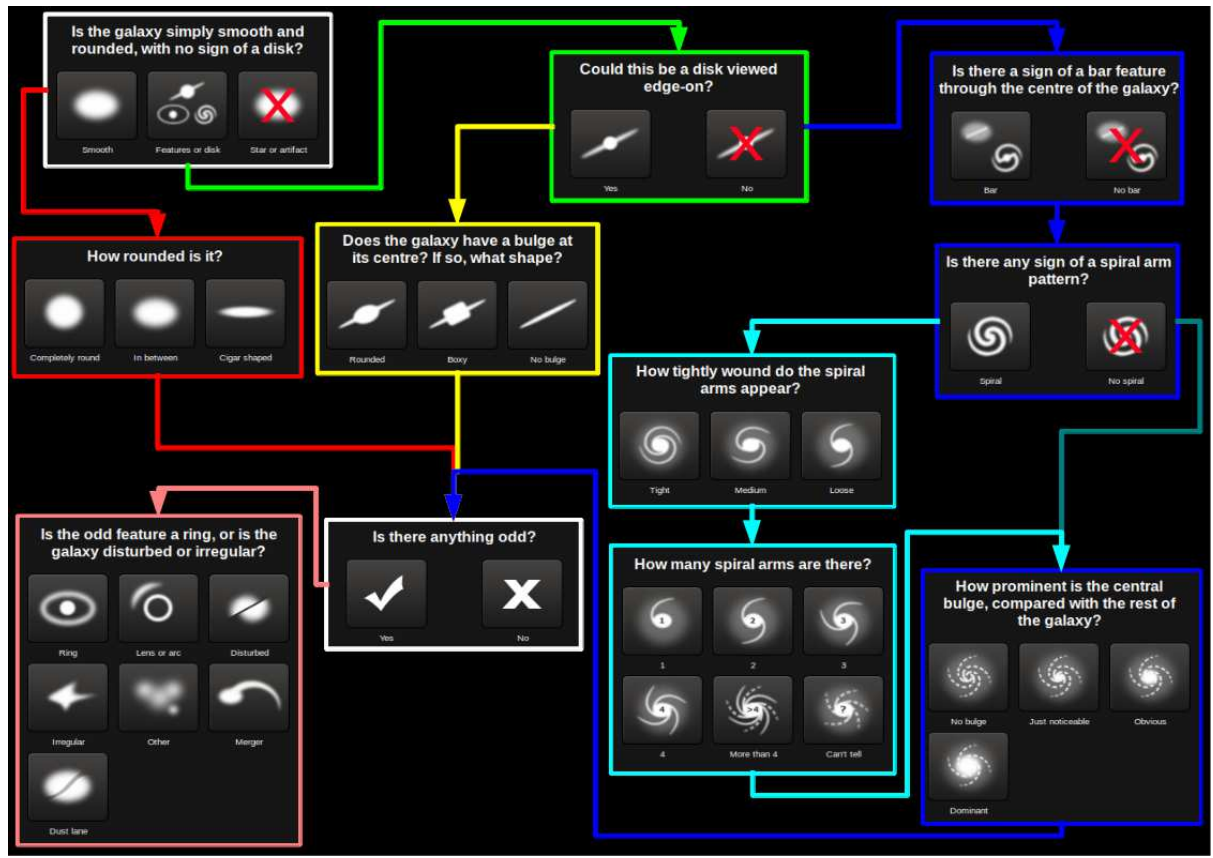}
\caption{The Galaxy Zoo 2 question tree}\label{gz2flow}
\end{figure*}

Figure \ref{gz2flow} presents the classification tree used for GZ2, including the actual button images shown to the GZ2 
participants.  These images attempt to symbolise the answer to each question.  Most of the participants have no formal astronomy training and, although there is a tutorial which presents real examples for each answer, it is likely that they rely on these images to a reasonable degree while making their classifications.

The GZ2 sample selection includes $273\,783$ galaxies from SDSS DR7 with $R_{90} > 3 \arcsec$ and $r_\rmn{Petro,AB} < 17.0$ and located in the contiguous North Galactic Cap region\footnote{GZ2 also included $\sim$30,000 galaxies from both the normal- and coadd-depth imaging of the SDSS Stripe 82 region, but these data are not used in this paper.}.  In this paper we require redshift information, and therefore restrict our sample to galaxies with spectra in DR7.  The overall spectroscopic completeness of this GZ2 sample is $86$ per cent.
As in \citet{mast2010}, we limit our sample of galaxies to $z<0.09$ to ensure sufficient resolution for spiral classifications and thereby reduce somewhat the redshift classification bias discussed in \citet{bamf2009}. 
A lower redshift limit of $z>0.01$ is imposed to avoid the de-blending of large nearby objects into multiple photometric objects by the SDSS pipeline, and ensure that redshift-derived distance moduli are sufficiently accurate. 
Applying these limits results in a sample of $148\,291$ galaxies.

We make use of the MPA-JHU DR7 median stellar mass measurements\footnote{this catalogue is publicly available from
http://www.mpa-garching.mpg.de/SDSS/DR7/Data/stellarmass.html}, which are based on fits to the SDSS photometry, following the methods of \citet{kauf2003} and \citet{sali2007}.

\section{Method}\label{method}

\subsection{Pair and control samples}

In this study we focus on galaxy pairs with stellar mass ratios between 1:1 and 4:1.  These would amount to major mergers, if the interaction were to proceed that far.
The candidate galaxy pairs in our parent sample are selected to have an absolute line of sight velocity difference of $\Delta V < 5000$ \kms and a projected physical separation (converted from an angular separation using the average redshift of the pair) of $\rp < 1000$ \hkpc.

To probe the effects of changing $\Delta V$ between pairs we divide our sample into $100$ \kms bins for 
$\Delta V < 500$ \kms, as well as two larger $\Delta V$ bins with $500< \Delta V < 1000$ \kms and 
$1000 < \Delta V < 5000$ \kms.
The $1000 < \Delta V < 5000$ \kms pairs serve as our control sample, as they should all be physically unassociated.  These are used to account for biases that result in a dependence of some morphological classifications on the projected 
separation of galaxies, irrespective of any true interaction.  They also indicate the typical level of morphological features present in the general galaxy population, for comparison with our pair samples.

Note that pairs with $\Delta V > 5000$ \kms have a significantly different luminosity distribution (skewed to more luminous galaxies) compared to the lower $\Delta V$ samples.  This is a result of $\Delta V$ approaching the velocity range of the survey, and hence the sample becoming dominated by more distant, intrinsically brighter galaxies. We judge the $1000 < \Delta V < 5000$ \kms range to be a good compromise between maintaining consistent luminosity selections, while maximising the control sample size and minimising its contamination by physically associated pairs.

\subsection{Morphology probabilities}\label{morphprob}

GZ2 participants are asked a series of questions for each image, with each answer determining the subsequent question, as depicted in Fig.~\ref{gz2flow}.  We record the individual answers provided and, after some down-weighting of the most inconsistent participants, use these to construct a catalogue. This catalogue gives the total number of times each galaxy was presented and, for each question, the fraction of votes for each possible answer, which we denote $f(A)$.  The $f(A)$ for all $A$ belonging to a single question must, of course, sum to unity.  To give a concrete example, consider that a given galaxy has been presented to 20 participants, 16 of which answered `Features or disk' to the initial question `Is the galaxy simply smooth and rounded, with no sign of a disk?'.  This galaxy thus has a vote fraction of $f({\emph{Features}}) = 0.8$.
%For brevity we unambiguously specify each combination of question and answer, and hence denote this as $f(\rmn{\emph{Features}}) = 0.8$.  
The 16 participants who answered \emph{Features} were then asked `Could this be a disk viewed edge-on?', with 12 answering `No', and thus $f(\rmn{\emph{Edge-on=No}}) = 0.75$.  All these 12 were subsequently asked `Is there a sign of a bar feature through the centre of the galaxy?', resulting in a split of $f(\rmn{\emph{Bar=Yes}}) = 0.5$ and $f(\rmn{\emph{Bar=No}}) = 0.5$.  Again, all 12 were then asked `Is there any sign of a spiral arm pattern?', with a `Yes' answer fraction of $f(\rmn{\emph{Spiral=Yes}}) = 0.75$.  The question tree continues, and we have only considered the route through the question tree taken by the majority, but this is sufficient for the reader to fully understand both the traversal of the question tree and the definition of $f(A)$.

The vote fraction $f(A)$ may, very roughly, be considered to represent probabilities regarding a galaxy's morphology.  The uncertainty expressed by these probabilities results from a combination of limited observational information, true morphologies which do not precisely align with the possible answers, and differing interpretations by each participant in judging the correspondence between the image and each answer.  In this case $f(A) = p(A | \bmath{M})$, the conditional probability of the galaxy having morphological feature $A$ given that it possesses the set of morphological features $\bmath{M}$, since each participant must have identified the galaxy as having the preceding features in the question tree,  $\bmath{M}$, in order to be asked about feature $A$.

The conditional probability compares the likelihood of a galaxy having feature $A$ against the alternative answers for a single question.  However, it does not necessarily give a good representation of the presence of a particular morphological feature.  An object may have a high $f(A)$, but still be unlikely to possess morphological attribute $A$.  To assess the overall likelihood of $A$, one may calculate an estimate for the joint probability of $A$ and $\bmath{M}$.  Formally this is,
\begin{align}
p(A \cap \bmath{M}) &= p(A | \bmath{M}) \, p(\bmath{M})\\
&= f(A) \prod_{\bmath{Q} \subset \bmath{M}}{\sum_{a \in \bmath{Q}}{f(a)}}\,,
\end{align}
where $\bmath{Q}$ are subsets of the set of answers $\bmath{M}$ partitioned by question, and the $a$ are the individual answers in $\bmath{Q}$.

As we do not distinguish between the different paths which lead to a given question (though this would be possible to do from the raw data), we only consider cases for which $\bmath{M}$ represents the sum of all possible paths to answer $A$.  In this case, we denote $p(A) = p(A \cap \bmath{M})$, where $p(A)$ is the probability of morphological feature $A$ together with any $\bmath{M}$ for which asking the question with answer $A$ is relevant, as defined by the question tree in Fig.~\ref{gz2flow}.  Remember that these probabilities only include the observational information available in GZ2, so do not necessarily equate to the true probability of a particular morphology feature.  For example, for a galaxy which has $f(\rmn{\emph{Features=Yes}}) < 1$ or $f(\rmn{\emph{Edge-on=No}}) < 1$, the value of $p(\rmn{\emph{Spiral=Yes}})$ does not take into account the unobservable (in terms of GZ2) presence of spiral arms in apparently smooth or edge-on galaxies.

For the example above,
\begin{align}
p(\rmn{\emph{Spiral=Yes}} \,|\, \rmn{\emph{Features}} \cap \rmn{\emph{Edge-on=No}})  &=\notag\\
f(\rmn{\emph{Spiral=Yes}}) &= 0.75\,,
\end{align}
whereas,
\begin{align}
p(\rmn{\emph{Spiral=Yes}}) =\qquad&\notag\\
p(\rmn{\emph{Spiral=Yes}} \cap \rmn{\emph{Features}} \cap \rmn{\emph{Edge-on=No}}) =\quad&\notag\\
f(\rmn{\emph{Spiral=Yes}}) f(\rmn{\emph{Edge-on=No}}) f(\rmn{\emph{Features}}) =&\ 0.45\,.
\end{align}
  This indicates that there is only a moderate probability that the object in question has visible spiral arms, although if one is willing to accept that it does have features and is not edge-on, then it probably does possess spiral arms.
Notice that the \emph{Bar} question has been implicitly omitted from this calculation as all its possible answers continue on to the \emph{Spiral} question, and hence it would contribute a factor of $\left[f(\rmn{\emph{Bar=Yes}}) + f(\rmn{\emph{Bar=No}})\right] = 1$.

Whether one works with $f(A)$ or $p(A)$ depends upon the question one is considering.  It is particularly useful to examine trends in terms of $f(A)$ itself, as this reflects the behaviour of a specific morphological feature, irrespective of other morphological variations.  However, due to the limited total number of times each object is considered, when $p(A)$ is low, $f(A)$ will be highly quantised and subject to high Poisson noise.  In this case, one can consider only objects for which $f(A)$ is meaningful, by imposing a minimum threshold on $p(\bmath{M})$, which we denote by $f(A \,|\, p(\bmath{M})\!>\!t)$, for some threshold $t$.

To study the dependence of $f(A)$ on projected separation ($\rp$) for a particular sample, we take the mean vote fraction $f(A)$ in each bin of $\rp$,
\begin{equation}\label{rpsum}
  f(\rp, A) = \frac{1}{N} \sum_{i=1}^N{f(A)_i}\;,
\end{equation}
where the sums are over the $N$ galaxies in each $\rp$ bin for that sample. This is done for each
of the answers, $A$, shown in Fig.~\ref{gz2flow}.

For many answers we find that the control sample displays a dependence on $\rp$.  Given their velocity separation, these trends cannot be due to any physical interactions within the pairs.  Instead they must arise from the apparent close projection of the galaxies, and may be considered to be a `projection bias', which will contaminate any signature of physical interaction in the lower $\Delta V$ pairs.  In order to remove this contamination, we remove the control sample trends versus $\rp$ from the observed trends for the other pair samples. 
If we regard the $f(\rp, A)$ as conditional probabilities, the control sample trend may be removed by,
\begin{equation}\label{subtract}
  F = (\fpair-\fcontrol) \,/\, (1-\fcontrol)\,,
\end{equation}
where $F$ is the conditional probability in the absence of the projection bias,
$\fpair$ is the measured conditional probability for a sample of physically associated pairs, and
$\fcontrol$ is the measured conditional probability for the control sample.
$F(\rp, A)$ therefore represents the conditional probability, $P(\rp, A | \bmath{M})$, of morphological feature $A$ being observed as a result of the galaxy being in a pair with separation $\rp$, given that the galaxy displays morphological features $\bmath{M}$.

The uncertainties on $F(\rp, A)$ are given by,
\begin{equation}\label{errors}
  \sigma_{F}^2 = \frac{(1 - \fpair)^2}{(1 - \fcontrol)^4}\; \sigma_{\fcontrol}^2 + \frac{1}{(1-\fcontrol)^2}\; \sigma_{\fpair}^2\;.
\end{equation}
where $\sigma_{\fcontrol}$ and $\sigma_{\fpair}$ are determined from the standard error of $\fcontrol$ and $\fpair$ in each $\rp$ bin.

Note that for physically associated pairs, with low $\Delta V$, the projected galaxy density increases as $\rp$ decreases, such that the number of galaxies in a given $\rp$ bin, and hence the signal-to-noise of $f(\rp, A)$ remains reasonably constant.  However, for unassociated pairs the projected galaxy density is constant as a function of $\rp$, and thus at small $\rp$ the signal-to-noise of $f(\rp, A)$ drops substantially.  Unfortunately, this limits the accuracy of the projection bias correction and translates into higher uncertainties in $F(\rp, A)$ at low $\rp$.

In order to quantify the strength and scale of the trends, we fit the $F(\rp, A)$ with the function,
\begin{equation}\label{fiteq}
  F(\rp) = a \exp(-\rp/b) + c\,,
\end{equation}
where $a$ is the size of the change in $F$ from large to small $r$, $b$ specifies the $r$ scale of the trend, and $c$ accounts for possible constant systematic offsets between the physically associated pairs and control sample.  We find that this simple empirical function is able to well represent most of the $F(\rp)$ trends we observe.  Ideally $c$ should be zero, and in any case, for $F(\rp)$ to represent a probability, $0 < c < (1-a)$.  However, systematic offsets of $c$ in either direction are possible due to small differences in the sample selections of physically selected pairs and the control sample.  These are difficult to avoid for different $\Delta V$ selections, but should not have an $\rp$ dependence, justifying the use of a simple constant to account for them.  Reassuringly, we find that $c$ is generally very close to zero, signifying that sample selection differences are indeed minimal.  The fitting method provides uncertainties on $a$, $b$ and $c$, which enables us to judge the significance of differences in the trends between samples and plot confidence intervals on the fitting functions.

\begin{figure*}
%\setlength{\tabcolsep}{0pt}
%\begin{tabular}{c c}
 % \includegraphics[width=60mm,angle=270]{fig2_oddyes.eps}&
 % \includegraphics[width=60mm,angle=270]{fig2_merger.eps}\\
 % \includegraphics[width=60mm,angle=270]{fig2_irregular.eps}&
 % \includegraphics[width=60mm,angle=270]{fig2_other.eps}\\
%\end{tabular}

\begin{tabular}{c c}
  \begin{overpic}[width=60mm,angle=270]{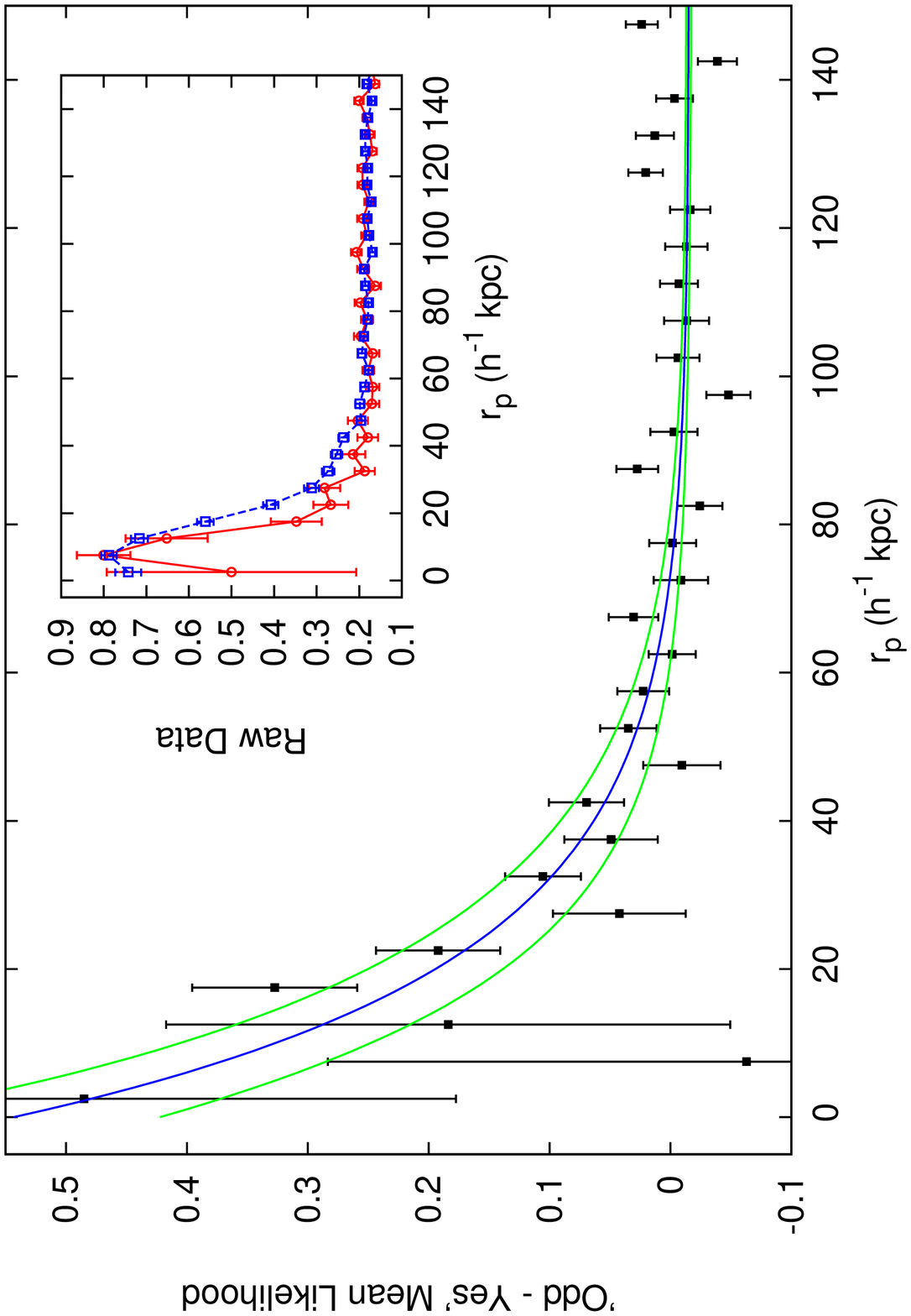}
	\put(55,145){\Large Odd=Yes}
  \end{overpic}&
  \begin{overpic}[width=60mm,angle=270]{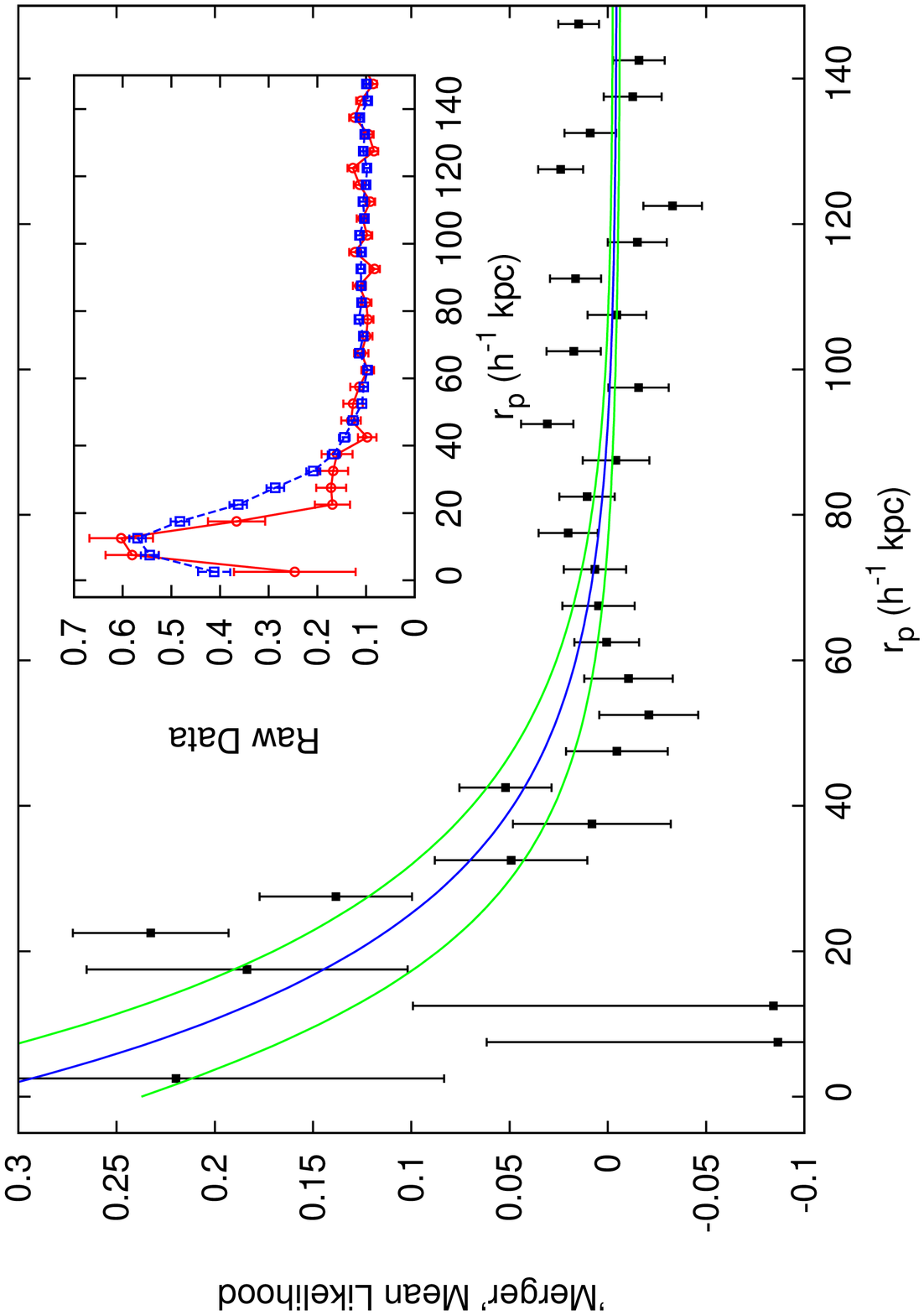}
	\put(55,145){\Large Merger}
  \end{overpic}\\
  \begin{overpic}[width=60mm,angle=270]{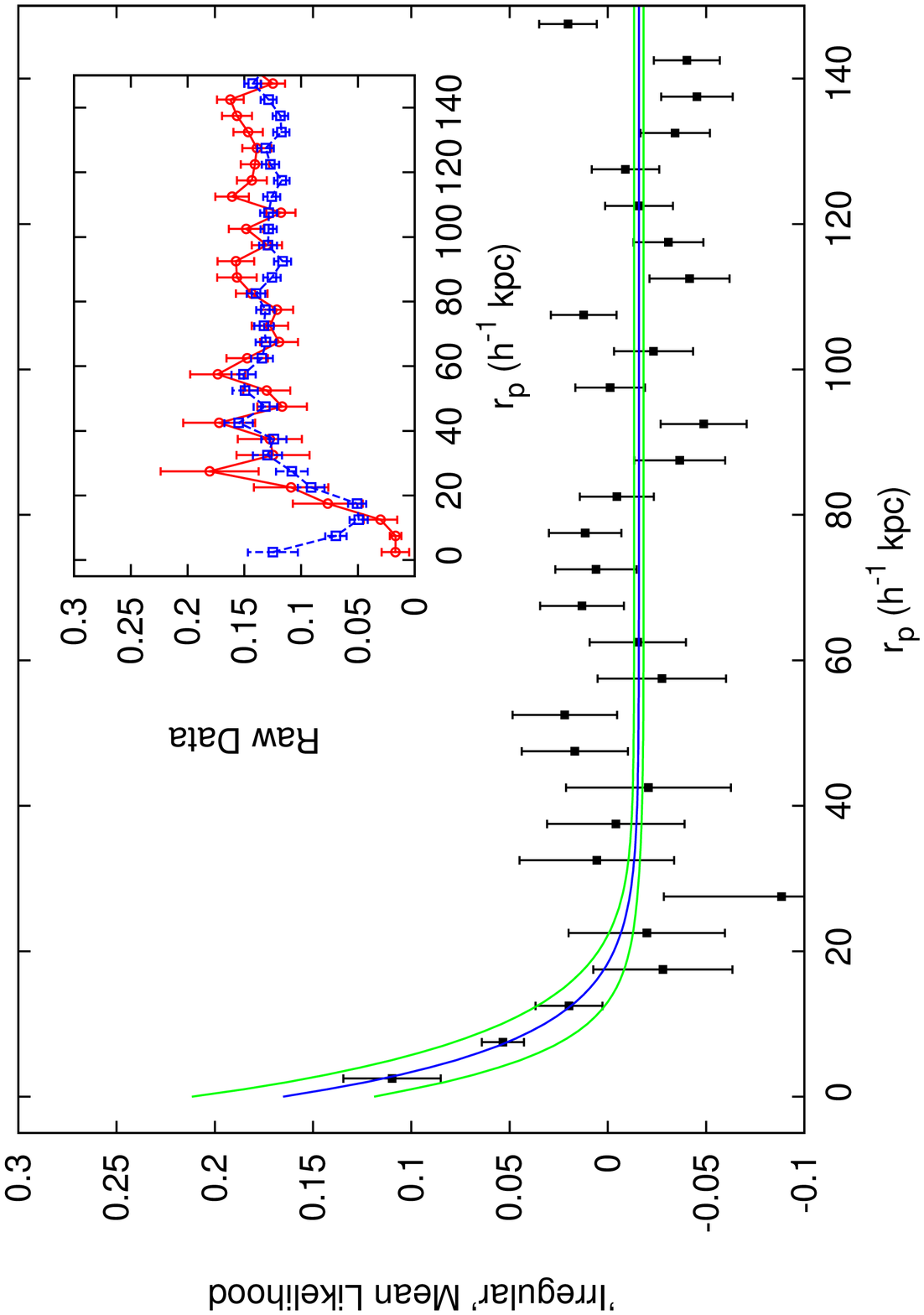}
	\put(55,145){\Large Irregular}
  \end{overpic}&
  \begin{overpic}[width=60mm,angle=270]{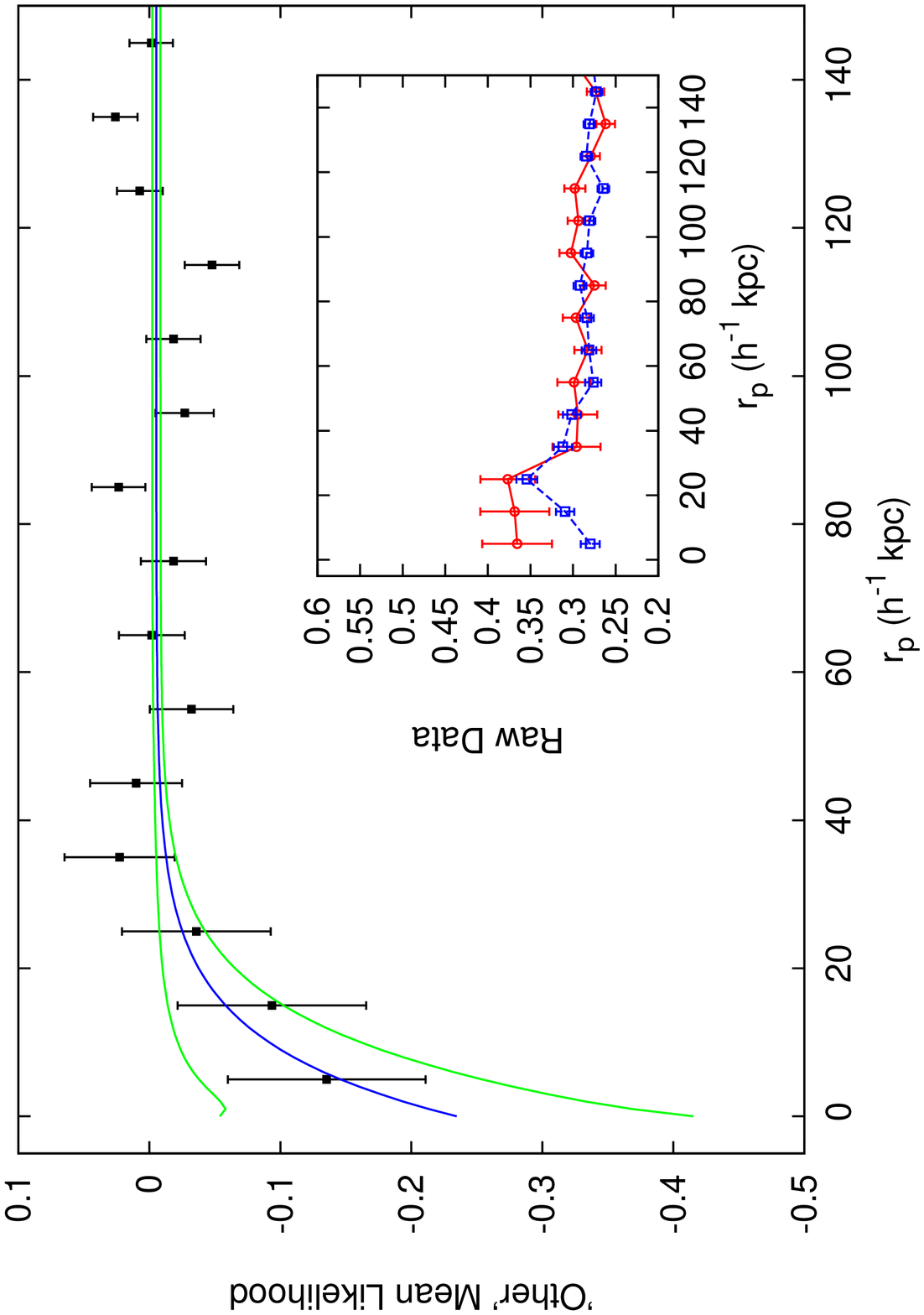}
	\put(55,40){\Large Other}
  \end{overpic}\\
\end{tabular}

  \caption{Mean vote fractions for the \emph{Odd=Yes} answer and subsequent `odd' categories which show trends versus $\rp$: \emph{Merger}, \emph{Irregular} and \emph{Other}.
The main plot in each panel shows the mean vote fraction of $\Delta V < 500$ \kms close pairs, corrected for projection bias, i.e. $F(\rp, A) = P(\rp, A | \bmath{M}$ (points with error bars), together with best fit (blue line) to these points using equation \ref{fiteq}, and the corresponding $1\sigma$ confidence region (green lines).  The insets plot $f(\rp, A)$ for the  $\Delta V < 500$ \kms close pairs (blue) and $1000 < \Delta V < 5000$ \kms control sample pairs (red), from which $F(\rp, A)$ is calculated.} \label{odd}
\end{figure*}

\subsection{Counting companions}\label{nccalc}

In Section \ref{interlopers} we study the number of close companions per galaxy as a function of projected separation, $\Nc(\rp)$, for a mass-limited sample.
%We only include galaxies with spectroscopic redshifts in this study and the overall spectroscopic completeness for the GZ2 sample,
%is 86 per cent, based on a SDSS limiting apparent magnitude of $r=17.0$.
%In the SDSS fiber collisions between the mechanisms 
%used to obtain spectra cause there to be angular spectroscopic incompleteness at angular separations less than 55 arcseconds.
%Spectra are only available at small angular separations in regions where individual SDSS fields overlap.
Following the methods of \citet{patt2000,patt2002},
weights are applied to account for pairs near the survey boundaries ($w_{b_2}$), pairs near the redshift boundaries ($w_{v_2}$), 
global spectroscopic incompleteness weights ($w_{s}$), as well as angular spectroscopic incompleteness weights to correct for fibre collisions at 
small angular separation ($w_{\theta_2}$).
%***********************************************
The uncertainties in these weights are determined primarily from uncertainties in the astrometric and redshift measurements.  Because the uncertainties in the pair statistics obtained in Section \ref{interlopers} are significantly larger than the astrometric and redshift uncertainties, uncertainties are not explicitly calculated for these weights.
%***********************************************

In Section \ref{interlopers} of this paper we will derive an additional weight ($w_\rmn{int}$) to account for the occurrence of interlopers: galaxies in close pairs (as judged by $\rp$ and $\Delta V$), but which are not truly interacting.  Previous studies typically ignore this, or estimate a constant value, whereas we will derive its dependence on $\rp$.

The total weight assigned to each galaxy is thus
\begin{equation}\label{weights}
w_{N_2} = w_{s}^{2}\, w_{\theta_2}\, w_{b_2}\, w_{v_2}\, w_\rmn{int}\;.
\end{equation}

The total number of companion galaxies of host galaxy $i$, with projected separation $\rp$, is given by summing these weights for all companions within a given projected separation,
\begin{equation}\label{ncieq}
N_{\rmn{c},i}(\rp) = \sum_{j = 1}^N {R(r_{\rmn{p},j}) \, w_{N_2, j}}\,,
\end{equation}
where $N$ is the total number of galaxies in the sample, and $R(r) = 1$ if $r$ is in the current $\rp$ bin, and $0$ otherwise.

The average number of close companions per galaxy, as a function of projected separation, is then calculated as the mean of the number of companions of each galaxy, weighted by the spectroscopic incompleteness,
\begin{equation}\label{nceq}
\Nc(\rp) = \frac{1}{N w_s} \sum_{i=1}^{N} N_{\rmn{c},i}(\rp)\;.
\end{equation}
For $\Nc(\rp) \ll 1$ this is equivalent to the fraction of galaxies with close companions.

%In sections \ref{oddclass} to \ref{deltav} we are interested
%in the fractions of galaxies which show a certain morphological feature at a given $\rp$, and need not apply spectroscopic weights.
%In section \ref{interlopers} we calculate the total number of interacting galaxies and thus apply the spectroscopic weights. 

%-Vmax weights to account for flux limited sample, apply vmax weight to faint member...

\section{Results}\label{results}

Following the method described in Section \ref{morphprob}, we have examined all of the GZ2 answers to
ascertain which are most relevant to studying galaxy interactions.  The questions which are of most interest 
for this paper are those regarding odd features, bars, the spiral arm winding tightness and the number of spiral arms.  As can be seen in Fig.~\ref{gz2flow}, for each object every participant is asked `Is there anything odd?' and, if they answer `Yes', asked to specify one odd feature from seven alternatives.  In order for a participant to be asked the questions regarding spiral arm number and winding tightness they must answer \emph{Features}, \emph{Edge-on=No} and \emph{Spiral=Yes} in the preceding set of questions. 

In sections \ref{oddclass}, \ref{armstails}, \ref{bars} and \ref{deltav}, we use the bias-corrected vote fractions, $F(\rp, A)$, corresponding to the probability of galaxy property $A$ occurring as a result of the galaxy being in a pair with separation $\rp$.  For example, $F(\rp, \rmn{\emph{1 Arm}}) = P(\rp, \rmn{\emph{1 Arm}} \,|\, \emph{Features} \cap \emph{Edge-on=No} \cap \emph{Spiral=Yes})$.  We use the full sample, without applying any thresholds for the preceding questions, in order to see what classification trends exist for that morphological feature in relation to pair separation irrespective of other morphological features.

In sections \ref{probpairs} and \ref{interlopers}, we select individual galaxies based
on their morphological features, and wish to minimize the impact of noisy $f(A)$ values for galaxies with low $p(A)$.  We therefore impose a threshold on $p(A)$ to select only objects for which asking the question with answer $A$ is likely to be appropriate.

\subsection{The Odd class}\label{oddclass}

\begin{figure}
  \begin{overpic}[width=60mm,angle=270]{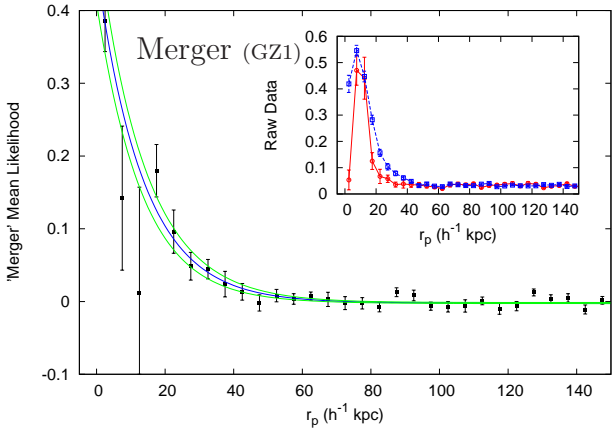}
	\put(55,145){\Large Merger \small (GZ1)}
  \end{overpic}\\

  \caption{Mean vote fractions for the GZ1 \emph{merger} answer.
The main plot shows the mean vote fraction of $\Delta V < 500$ \kms close pairs, corrected for projection bias, i.e. $F(\rp, A) = P(\rp, A | \bmath{M}$ (points with error bars), together with best fit (blue line) to these points using equation \ref{fiteq}, and the corresponding $1\sigma$ confidence region (green lines).  The insets plot $f(\rp, A)$ for the  $\Delta V < 500$ \kms close pairs (blue) and $1000 < \Delta V < 5000$ \kms control sample pairs (red), from which $F(\rp, A)$ is calculated.}\label{gz1merger}
\end{figure}

In Fig.~\ref{odd} we study the trends in `Odd' GZ2 morphologies as a function of pair separation.  The main panels plot the mean vote fraction, $F(\rp, A)$, of $\Delta V < 500$ \kms close pairs, correcting for projection bias effects using the $1000 < \Delta V < 5000$ \kms control sample, as outlined in Section \ref{morphprob}.  This may be interpreted as the probability, $P(\rp, A | \bmath{M}$, of a galaxy in this sample displaying the specified morphological attribute, purely as a result of being in a pair with separation $\rp$.  The raw vote fractions, $f(\rp, A)$, from which the corrected quantities are determined, are also shown in the inset panels.

Considering first the \emph{Odd=Yes} answer itself, top left in Fig.~ \ref{odd}, we see from the inset plot that the raw vote fractions, $f(\rp, \rmn{\emph{Odd=Yes}})$ increase strongly with decreasing $\rp$ for both the $\Delta V < 500$ \kms close pairs and the control sample.  As the pairs in the control sample are physically unassociated, this indicates that GZ2 participants identified some galaxies as `Odd' simply as a result of their apparent separation.  This projection bias strongly contaminates the $f(\rp, \rmn{\emph{Odd=Yes}})$ results of the $\Delta V < 500$ \kms close pair sample.  However, there are offsets between the $f(\rp, \rmn{\emph{Odd=Yes}})$, such that the $\Delta V < 500$ \kms close pairs are more likely to be marked `Odd' than the control sample, particularly at $20 \la \rp \la 40$ \hkpc.

Indeed, when we correct for the projection bias, we see clear evidence for a trend.  GZ2 participants were more likely to identify an object as `Odd' if it is in a $\Delta V < 500$ \kms close pair with small projected separation.  The probability of a galaxy being labelled `Odd', as a result of being in a pair, $P(\rmn{\emph{Odd=Yes}})$, varies from zero for $\rp \ga 80$ \hkpc to $\sim 0.5$ for the smallest projected separations.  The empirical function defined in Eqn.~\ref{fiteq} does a good job of representing this trend.

If a participant answered \emph{Odd=Yes}, then they were then asked to identify the odd feature more precisely by selecting one option from a variety of possibilities.  The remaining panels of Fig.~\ref{odd} plot $F(\rp, A) = P(\rp, A \,|\, \rmn{\emph{Odd=Yes}})$ for three of these options.   From the top right panel of Fig.~\ref{odd} we see that the \emph{Merger} answer mimics the behaviour of \emph{Odd=Yes}, although note that we are plotting the conditional probability given that the object does display an `Odd' feature, and hence the \emph{Odd=Yes} behaviour itself is not included in this quantity.

We see, from the raw vote fractions, that the \emph{Merger} answer is strongly dependent on the apparent separation of galaxy pairs.  Galaxies are often marked as a `Merger' because they appear close together in the image, even when there are no other signs of interaction.  Despite this, there is a clear enhancement of \emph{Merger} features at $20 \la \rp \la 40$ \hkpc for $\Delta V < 500$ \kms close pairs over that seen for the control sample.  In the projection bias corrected $F(\rp, \rmn{\emph{Merger}})$ we also see that physically interacting low $\Delta V$ pairs do have an additional probability of being identified as a merger, although it is a noisy function of $\rp$.

The \emph{Merger} answer displays cross talk with the other `Odd' categories.
As $f(\rp, \rmn{\emph{Merger}})$ increases with decreasing $\rp$, the vote fractions of most of the alternative answers
(i.e. \emph{Ring}, \emph{Arc}, \emph{Disturbed}, \emph{Irregular}, \emph{Dust Lane}) decrease, for both the low $\Delta V$ and control sample pairs.  
However, with the projection bias accounted for, at the smallest separations ($\rp< 10$ \hkpc) $F(\rp, \rmn{\emph{Irregular}})$ and $F(\rp, \rmn{\emph{Disturbed}})$ increase, possibly at the expense of $F(\rp, \rmn{\emph{Merger}})$.  This is consistent with users preferentially classifying separated pairs as `Mergers' and interpreting overlapping pairs as single `Irregular' or `Disturbed' objects.  
An enhancement of GZ2 \emph{Dust Lane} features in merging early-types has already been discussed by \citet{shab2011}, although we only see a tentative indication of this with our method.

Interestingly, the \emph{Other} category displays a contrasting behaviour, showing a decreasing $F(\rp, \rmn{\emph{Other}})$ with decreasing $\rp$.  This appears to be a result of other, more specific, categories being favoured for pairs with a real physical association.   Looking at the \emph{Other} button image in Fig.~\ref{gz2flow} it is clear why some projected close pairs are given this classification, especially non-interacting early type galaxies.

All these results have important implications for the use of the GZ2 `Odd' classifications.  In particular, they imply that for physically interacting close pairs, the likelihood
of being classified as a merger actually begins to decreases at the smallest separations as galaxies begin to be classified more as irregular or disturbed.  Given the large amount of projection bias and the cross-talk between categories, it is difficult to use any of the `Odd' categories alone as indicators of pair interactions.  It seems that using the more general \emph{Odd=Yes} provides the most straightforward signal of interacting pairs.

For comparison, the GZ1 \emph{Merger} class is plotted in Fig.~\ref{gz1merger}. Since GZ2 uses a subset of the GZ1 sample, GZ1 classifications are available for all of the galaxies in our sample and the same galaxies are included in Figs.~\ref{odd} and \ref{gz1merger}. The GZ1 and GZ2 \emph{Merger} classifications display the same behaviour with decreasing projected separation, with both showing a dip at smallest separations ($\rp< 10$ \hkpc). For GZ2 many of these \emph{Merger} votes are exchanged for the \emph{Irregular} and \emph{Disturbed} categories, whereas for GZ1 it appears that the votes go to \emph{Spiral-CW} and \emph{Spiral-ACW} classifications, perhaps due to the formation of tidal arms.  It is possible that similar effects have been present, but gone unnoticed, in previous work using more traditional classification schemes.
Despite the presence of a significant signal for the control sample, in most separation-bins the low-$\Delta V$ pairs display higher \emph{Merger} vote fractions than high-$\Delta V$ pairs, and hence a sensible correction for this projection bias may be applied.

The parameters of the fitting functions, quantifying the amplitude and scale of the trends shown in Fig.~\ref{odd} and \ref{gz1merger} are provided in Table \ref{fitparam}.

\subsection{Spiral arms and tidal tails}\label{armstails}

\begin{figure*}
%\begin{tabular}{c c}
 % \includegraphics[width=60mm,angle=270]{fig4_spiral.eps}&
 % \includegraphics[width=60mm,angle=270]{fig4_2arm.eps}\\
 % \includegraphics[width=60mm,angle=270]{fig4_loose.eps}&
 % \includegraphics[width=60mm,angle=270]{fig4_1arm.eps}\\
%\end{tabular}
\begin{tabular}{c c}
  \begin{overpic}[width=60mm,angle=270]{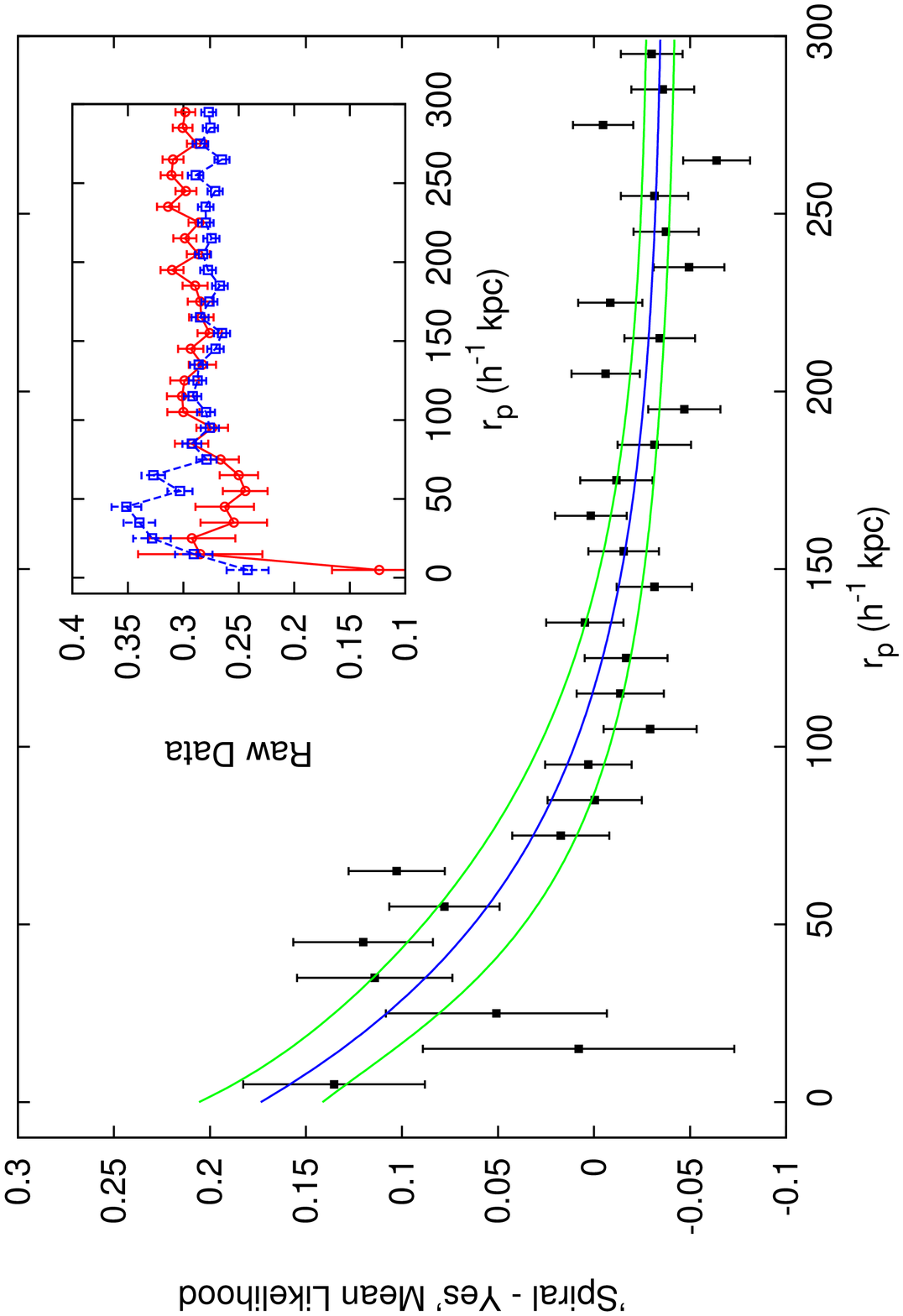}
	\put(55,145){\Large Spiral=Yes}
  \end{overpic}&
  \begin{overpic}[width=60mm,angle=270]{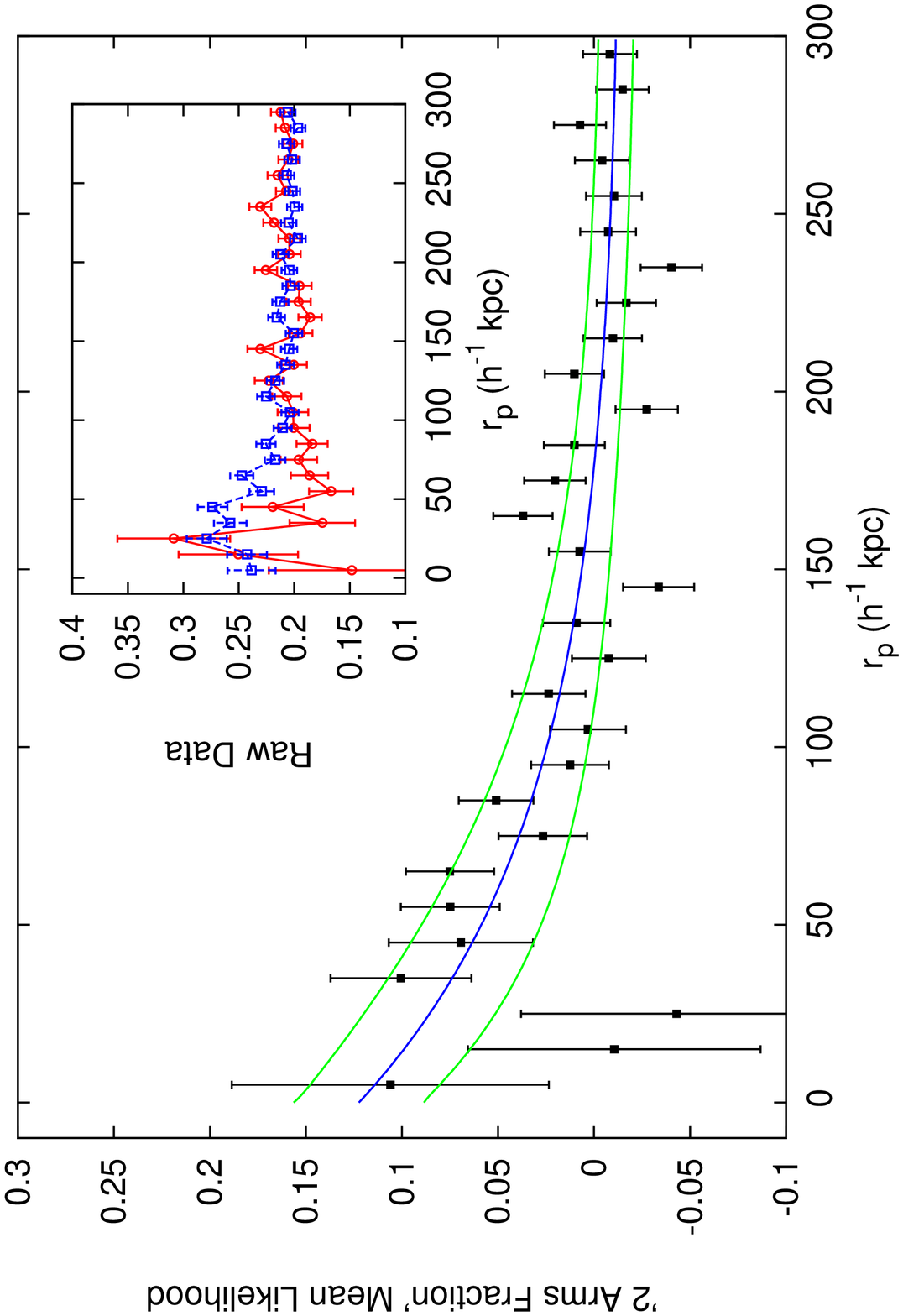}
	\put(55,145){\Large 2 Arms}
  \end{overpic}\\
  \begin{overpic}[width=60mm,angle=270]{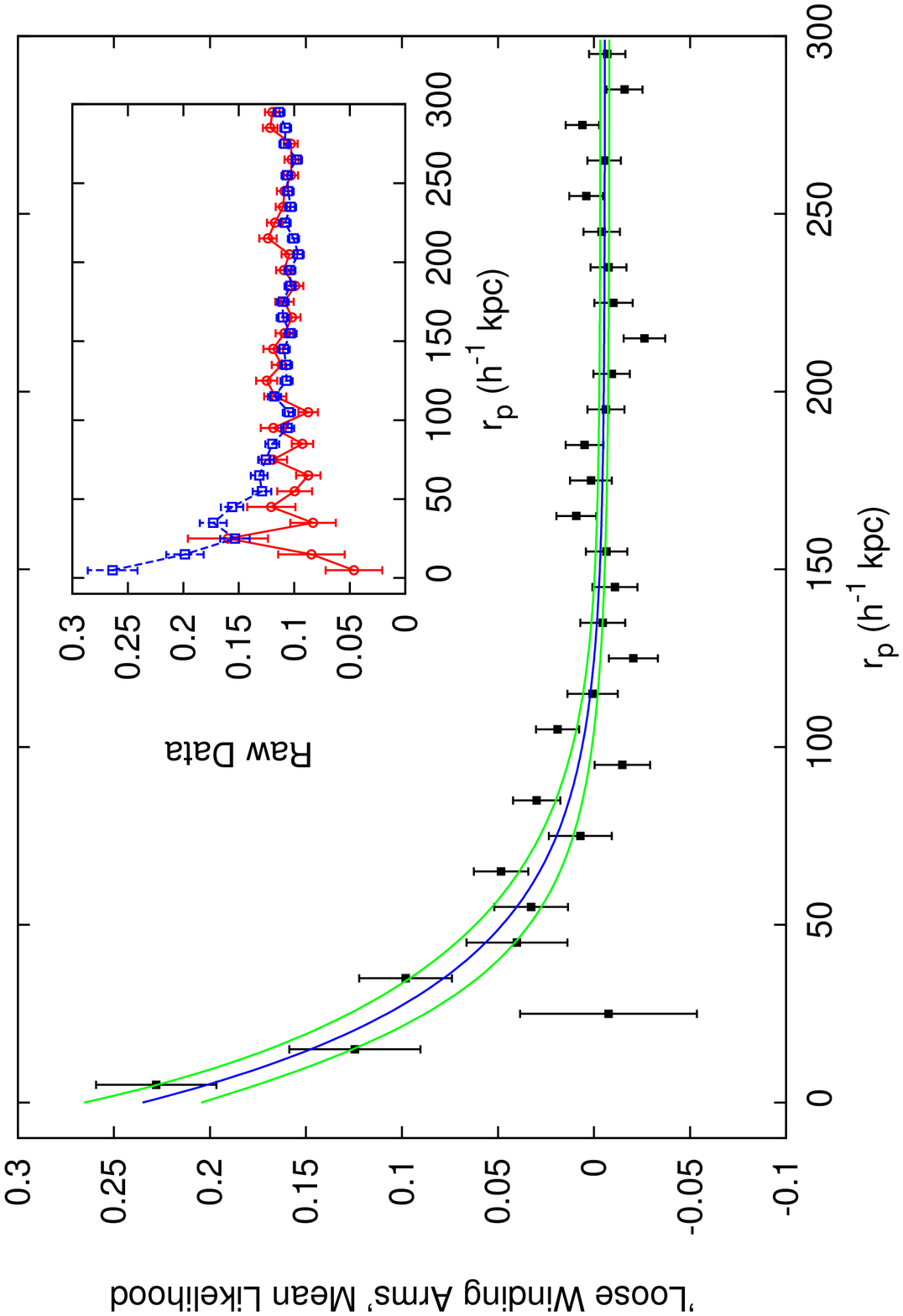}
	\put(55,145){\Large Loose Arms}
  \end{overpic}&
  \begin{overpic}[width=60mm,angle=270]{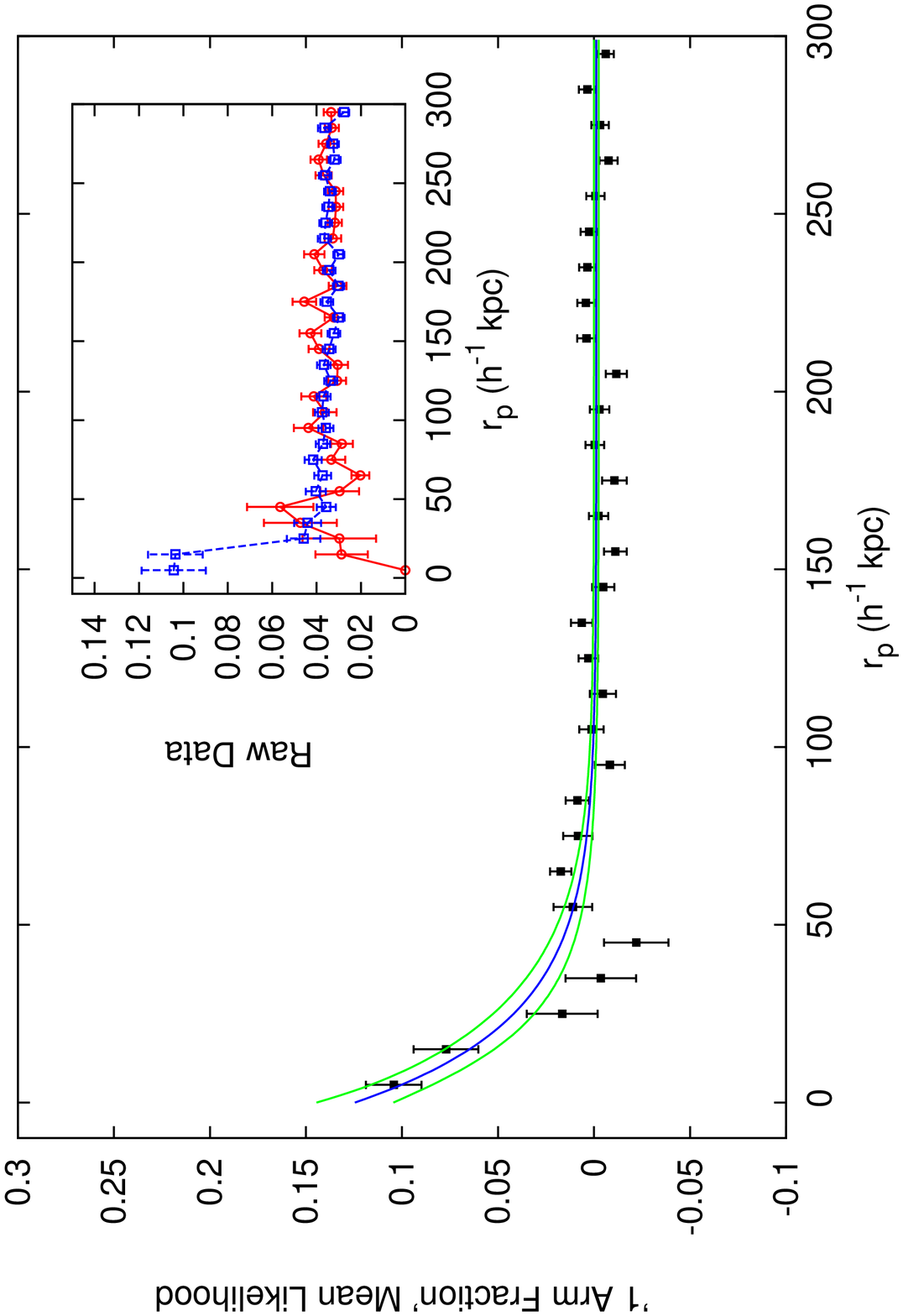}
	\put(55,145){\Large 1 Arm}
  \end{overpic}\\
\end{tabular}
\caption{Mean vote fractions for the \emph{Spiral=Yes} answer and answers to the subsequent `arm number' and `arm tightness' questions which show trends versus $\rp$: \emph{1 Arm}, \emph{2 Arms} and \emph{Loose Winding Arms}.
The main plot in each panel shows the mean vote fraction of $\Delta V < 500$ \kms close pairs, corrected for projection bias, i.e. $F(\rp, A) = P(\rp, A | \bmath{M}$ (points with error bars), together with best fit (blue line) to these points using equation \ref{fiteq}, and the corresponding $1\sigma$ confidence region (green lines).  The insets plot $f(\rp, A)$ for the  $\Delta V < 500$ \kms close pairs (blue) and $1000 < \Delta V < 5000$ \kms control sample pairs (red), from which $F(\rp, A)$ is calculated.}\label{tidal}
\end{figure*}

Although the `Odd' GZ2 questions were targeted directly at identifying interactions, there are other morphological indicators available.  Tidal interactions frequently induce or enhance spiral arm and tidal tail features.  Indeed, we find that a number of the GZ2 answers regarding spiral features display a clear dependence on pair separation.

Fig.~\ref{tidal} presents the trends in `Spiral' GZ2 morphological features as a function of pair separation.  As in Fig.~\ref{odd}, the main panels plot the mean vote fraction, $F(\rp, A)$, of $\Delta V < 500$ \kms close pairs, after correcting for projection bias effects using the $1000 < \Delta V < 5000$ \kms control sample (see \ref{morphprob}).  This quantity represents the probability, $P(\rp, A | \bmath{M})$, of a galaxy in this sample displaying the specified morphological attribute, purely as a result of being in a pair with separation $\rp$, provided it is meaningful to ask about that feature.  The raw vote fractions, $f(\rp, A)$, from which the corrected quantities are determined, are also shown in the inset panels.  Although we still see some trends with $\rp$ in the control sample, the projection biases are much less severe than those for the `Odd' features considered above.

We first consider the probability that galaxies will be classified as displaying spiral features, given that they display any features and are not edge-on disks, $F(\rp, \rmn{\emph{Spiral=Yes}}) = P(\rp, \rmn{\emph{Spiral=Yes}} \,|\, \rmn{\emph{Features}} \cap \rmn{\emph{Edge-on=No}})$. 
We find that this increases significantly with
decreasing projected separation, beginning around $\rp \la 100$ \hkpc.  We see that the trends in 
$F(\rmn{\emph{Spiral=Yes}})$ and $F(\rmn{\emph{2 Arms}})$ are similar, indicating that the increase in the probability of close pairs presenting spiral features is accompanied by an enhancement in the probability of those spiral features being two-armed.  This increase in the probability of \emph{2 Arms} is at the expense of the probability that the number of arms cannot be discerned.  Together, these indicate a general strengthening of two-arm spiral patterns in close pairs on an exponential scale of $\sim 70$ \hkpc.  This is consistent with observations and simulations which see an enhancement in spiral arm strength in interacting systems, often accompanied by an increase in star formation activity (e.g., \citealt{saud1987}).

We can also consider the occurrence of different numbers of spiral arms.  At small separations ($\rp \la 20$ \hkpc) the probability of a galaxy in a low $\Delta V$ pair displaying a single spiral arm, $F(\rmn{\emph{1 Arm}})$ increases sharply.  There is perhaps a suggestion that $F(\rp, \rmn{\emph{2 Arms}})$ decreases somewhat as \emph{1 Arm} increases at small $\rp$. The other answers to the
`How many spiral arms are there?' question show no significant change with decreasing pair separation.

Finally we examine how the winding tightness of any spiral pattern depends on pair separation.  In Fig.~\ref{tidal} we see that the probability of a galaxy being classified as having \emph{Loose Winding Arms} (hereafter \emph{LWA}) increases
for $\rp < 100$ \hkpc.  On the other hand, the \emph{Medium Winding Arms} and \emph{Tight Winding Arms} classes show no significant change with $\rp$. 
Although it has been known for some time that strong tidal interactions between galaxies can produce either 1 or 2 loosely wound tidal arms, depending on the orbital parameters (e.g., \citealt{thom1989,howa1993,barn2009,barn2011}), this is the first study to determine the observability of these features as a function of separation.

The \emph{1 Arm} and \emph{LWA} features appear to be particularly robust indicators of pair interaction, as the control samples display opposing behaviour to the physically associated pairs.  In the absence of a true interaction, galaxies in projected pairs are actually less likely to be classified as having \emph{1 Arm} or \emph{LWA}.  Given the limited impact of projection bias effects for these quantities, we could in principle use the low $\Delta V$ pair trends in arm winding and arm number directly, without the need for remove the $\rp$ dependence of the control sample.  However, performing this correction gives us a more reliable and quantitative measurement of the trends.  A price of this correction is noise added due to the limited size of the control sample.

It was originally envisaged that these questions would primarily provide information regarding the usual spiral arms, but it seems clear that in the case of close pairs they are revealing additional information.  Observations and simulations of strong tidal galaxy interactions frequently show extended, asymmetric tails (see the discussion and references in Section \ref{intro}).  We therefore interpret the trends in $F(\rp, \rmn{\emph{1 Arm}})$ and $F(\rp, \rmn{\emph{LWA}})$ as unambiguous signatures of tidal tails caused by interactions between close pairs.

We also see an increase in the \emph{Features} and \emph{Edge-on=No} answers with decreasing pair separation. The increase in \emph{Features} appears to be a result of galaxies that would have otherwise appeared smooth and featureless gaining enhanced spiral arms or tidal features in close pairs. Similarly, spiral galaxies which are being tidally disturbed
may develop warped disks, and therefore be less likely to be classified as \emph{Edge-on=Yes}.  This could have the additional effect of preserving the visibility of the spiral arms, despite the high inclination.  Also, at least in some cases, it appears that participants may interpret a warped edge-on disc as \emph{LWA}.

The amplitude and scale of the trends shown in Fig.~\ref{tidal}, in terms of the parameters of the best fit of Eqn.~\ref{fiteq}, are provided in Table \ref{fitparam}, along with the corresponding values for \emph{Features} and \emph{Edge-on=No}.

\begin{table}
 \centering
 %\begin{minipage}{140mm}
  
  \begin{tabular}{c c c c c}
  \hline Answer  & $a$ & $b$ (\hkpc) & $c$\\
 \hline
\emph{Odd=Yes} & 0.54$\pm$0.15 & 21$\pm$3 & -0.015$\pm$0.002\\
\emph{Irregular} &  0.18$\pm$0.05 & 8$\pm$2 & -0.016$\pm$0.002\\
\emph{Disturbed} & 0.10$\pm$0.02 & 6$\pm$2 & 0.005$\pm$0.002\\
\emph{Other} & -0.23$\pm$0.19 & 10$\pm$9 & -0.005$\pm$0.003\\
\emph{Merger} &  0.31$\pm$0.10 & 21$\pm$4 & -0.004$\pm$0.002\\
\emph{GZ1 Merger} & 0.46$\pm$0.05 & 14$\pm$1 & -0.002$\pm$0.001\\
%Other (5 kpc bin) &  -1.17$\pm$2.28 & 3.1$\pm$2.6 & -0.005$\pm$0.003\\
 \hline
\emph{Features} & 0.22$\pm$0.04 & 56$\pm$13 & -0.033$\pm$0.006\\
\emph{Edge-on=No} & 0.55$\pm$0.13 & 39$\pm$11 & 0.005$\pm$0.014\\
%\emph{Bar=No} & 0.59$\pm$0.17 & 29$\pm$9 & 0.017$\pm$0.012\\
%\emph{Bar=No} & 0.44$\pm$0.20 & 31$\pm$13 & 0.013$\pm$0.012\\
\emph{Bar=No} & 0.46$\pm$0.20 & 28$\pm$12 & 0.013$\pm$0.012\\
\emph{Spiral=Yes} & 0.21$\pm$0.04 & 66$\pm$16 & -0.036$\pm$0.008\\
\emph{Loose Winding Arms} & 0.24$\pm$0.03 & 33$\pm$5 & -0.006$\pm$0.003\\
\emph{1 Arm} & 0.13$\pm$0.02 & 23$\pm$4 & -0.001$\pm$0.001\\
\emph{2 Arms} & 0.14$\pm$0.04 & 79$\pm$32 & -0.014$\pm$0.009\\
\hline
\end{tabular}\caption{Best fit results to the plots of $F(\rp, A)$ in Figs. \ref{odd}, \ref{gz1merger}, \ref{tidal} and \ref{barno} using Eqn.~\ref{fiteq}.
\emph{Edge-on=No} and \emph{Bar=No} are fits to trends with 20 \hkpc bins,
\emph{Spiral}, \emph{Loose Winding Arms}, \emph{1 Arm}, \emph{2 Arms} and \emph{Other} use 10 \hkpc bins, while
\emph{Odd}, \emph{Merger}, \emph{Irregular}, \emph{Disturbed} and \emph{Features} use 5 \hkpc bins.}\label{fitparam}
%\end{minipage}
\end{table}

\subsection{Barred galaxies}\label{bars}

The identification and properties of barred galaxies in Galaxy Zoo is studied in detail by \citet{mast2011,mast2012} and \citet{hoyl2011}. The dependence of GZ2 bars on environment is presented in \citet{skib2011}.  Here we continue our focus on the relationship between morphology and close pair interactions.

\begin{figure}
  \begin{overpic}[width=60mm,angle=270]{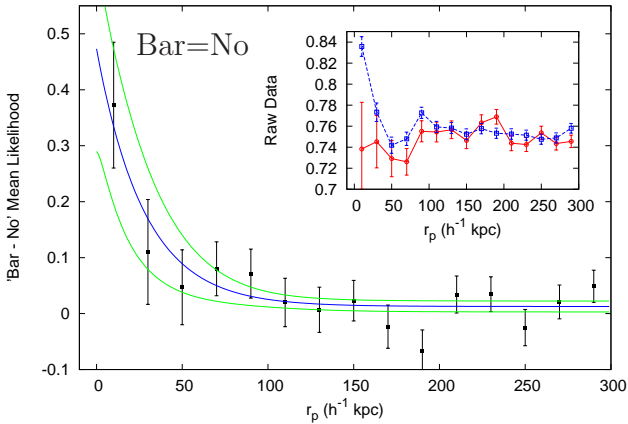}
	\put(55,145){\Large Bar=No}
  \end{overpic}
  \caption{Mean vote fractions for the \emph{Bar=No} answer.
The main plot shows the mean vote fraction of $\Delta V < 500$ \kms close pairs, corrected for projection bias, i.e. $F(\rp, A) = P(\rp, A | \bmath{M}$ (points with error bars), together with best fit (blue line) to these points using equation \ref{fiteq}, and the corresponding $1\sigma$ confidence region (green lines).  The insets plot $f(\rp, A)$ for the  $\Delta V < 500$ \kms close pairs (blue) and $1000 < \Delta V < 5000$ \kms control sample pairs (red), from which $F(\rp, A)$ is calculated.}\label{barno}
\end{figure}

In Fig.~\ref{barno} the corrected mean vote fraction for the \emph{Bar=No} answer, $F(\rp, Bar=No)$, is plotted versus projected separation. This trend was found to be rather noisy so we use $20$ \hkpc $\rp$ bins to more clearly represent the data. There appears to be a increase in $F(\rp, Bar=No)$ for $\rp < 20$ \hkpc in the low-$\Delta V$ pairs, while the high-$\Delta V$ control sample pairs show no significant change with projected separation.

This result indicates that bars are suppressed, rather than triggered, by strong tidal interactions.  It is possible that part of the observed trend is a result of bars becoming less noticeable in interacting pairs due to increased star formation and looser spiral arms, but even so there cannot be a strong enhancement of bars in our interacting pairs.  This is in agreement with the preliminary findings of \citet{mend2011} who compared isolated and paired galaxies and found the bar fraction to be $\sim 43$ per cent for isolated galaxies, but only $\sim 20$ per cent for pairs (where bars are identified in ellipse fits to the isophotes).  Similarly, \citet{lee2012} find the fraction of visually classified strong bars decreases at small pair separations.

At first this seems contrary to the expectation that bar features are excited by gravitational interactions, as indicated by many simulation studies (e.g., \citealt{nogu1988, moor1996, moor1998,roma2008}), although in simulations including gas, tidal interactions appear less able to generate bars (e.g., \citealt{bere2004}).  There have also been observational indications of bar enhancement in dense environments and interacting systems (e.g., \citealt{elme1990}).  However, disk galaxies in denser environments are more massive and possess redder colours and earlier morphologies, all of which also correlate with the presence of bars \citep{mast2011}.  In a detailed study of this issue, \citet{skib2011} found a substantial enhancement in the appearance of bars in denser environments, but concluded that the majority of the effect was attributable to the dependence of colour and stellar mass on environment.  However, a significant correlation remains, which \citeauthor{skib2011} argue indicates that minor mergers and tidal interactions increase the appearance of bars by triggering disc instabilities. They also find the bar-environment correlation to decrease at small separations, becoming statistically insignificant at $\rp<150$ \hkpc.  While not identical to our close pair results, this is nevertheless consistent.  Simulations also demonstrate the suppression or destruction of bars in strong tidal interactions \citep{bere2003}.  However, most simulation work focuses on the final result of pair interactions, rather than the evolution of morphological features over the course of the interaction.  In the low-mass ratio pairs we study in this paper, the final result of the interaction may often be the destruction of any bars along with their entire disk.  Theoretically it seems that destruction of bars at earlier stages of pair interactions is feasible, and the observations, in this paper and by others, appear to indicate this.  The emerging picture is therefore that moderate interactions, with high-mass ratios (i.e., minor mergers and harassment), promote bar formation, but that stronger interactions suppress the appearance of bars.

\subsection{Dependence of tidal effects on line-of-sight velocity difference}\label{deltav}

\begin{figure}
  \begin{overpic}[width=60mm,angle=270]{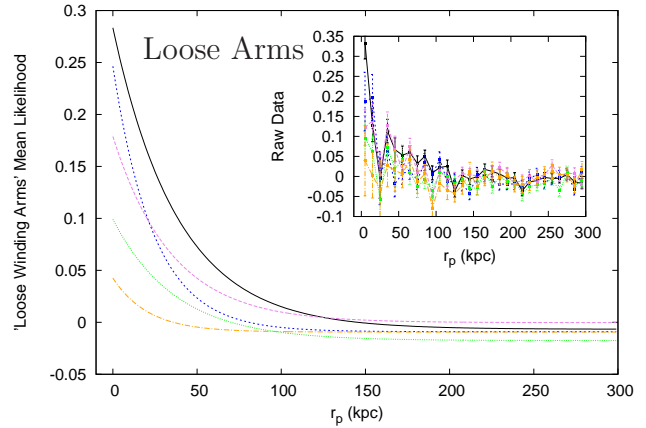}
	\put(55,145){\Large Loose Arms}
  \end{overpic}
  \caption{The trend in the probability of observing \emph{Loose Winding Arms} with projected separation, $F(\rp, \rmn{\emph{LWA}})$, for pair samples selected with different ranges in $\Delta V$: $0<\Delta V<100$ \kms \emph{(black)}, $100<\Delta V<200$ \kms \emph{(purple)}, $200<\Delta V<300$ \kms \emph{(blue)}, $300<\Delta V<400$ \kms \emph{(green)}, and $400<\Delta V<500$ \kms \emph{(yellow)}. The low $\Delta V$ pairs clearly show stronger signs of interaction than the high $\Delta V$ pairs.}\label{vlim}
\end{figure}

So far we have been considering the $\rp$ dependence for a fixed sample of pairs with $\Delta V < 500$ \kms, versus a control sample with $1000 < \Delta V < 5000$ \kms.  However, we would also expect a relationship between physical separation, and hence interaction strength, and $\Delta V$.  This is explored in Fig.~\ref{vlim}, in which we plot $F(\rp, \rmn{\emph{LWA}})$ for pairs in several bins of $\Delta V$.   Again, we fit the data with Eqn.~\ref{fiteq} in an attempt to quantify the trends.  Due to the limited statistics, these plots are relatively noisy and the fit parameters are sometimes quite uncertain.  Nevertheless, some trends are clearly seen in both the raw data and the fits.

Signs of interaction appear strong for the smallest $\Delta V$ pairs, and weaken with increasing $\Delta V$.  We see almost no signs of interaction for $\Delta V > 500$ \kms. In Table \ref{dvfits} we see that for $\Delta V > 300$ \kms the fitting parameter $a$ (representing the excess \emph{LWA} likelihood at $\rp = 0$ \hkpc) decreases sharply.
As $\Delta V$ decreases, there is a hint that pairs show signs of tidal interactions further out.  This would be consistent with our expectations: pairs with low $\Delta V$ must be predominantly interacting in the plane of the sky, whereas at higher $\Delta V$, pairs will be increasingly interacting along the line-of-sight.

In future work we intend to treat $\Delta V$ and $\rp$ on similar terms, quantifying the dependence of observed morphological features as a function of both quantities.

%\begin{table}
%\centering
%\begin{tabular}{c c c c c}
%\hline  $\Delta V$ (\kms) & $a$ & $b$ (\hkpc) & $c$\\
%\hline
%$0<\Delta V < 100$ & 0.19$\pm$0.03 & 42$\pm$6 & 0.104$\pm$0.003 \\
%$100<\Delta V < 200$ &0.13$\pm$0.04 & 31$\pm$8 & 0.109$\pm$0.003\\
%$200<\Delta V < 300$ &0.22$\pm$0.08 & 19$\pm$6 & 0.102$\pm$0.003\\
%$300<\Delta V < 400$ &0.06$\pm$0.04 & 43$\pm$29 & 0.094$\pm$0.004\\
%$400<\Delta V < 500$ &0.00$\pm$0.10 & 19$\pm$401 & 0.100$\pm$0.004\\ 
%\hline
%\end{tabular}\caption{Fitting parameters to the curves in Figure \ref{vlim}...}\label{dvfits}
%\end{table}

\begin{table}
\centering
\begin{tabular}{c c c c c}
\hline  $\Delta V$ (\kms) & $a$ & $b$ (\hkpc) & $c$\\
\hline
$0<\Delta V < 100$ & 0.29$\pm$0.04 & 39$\pm$5 & -0.007$\pm$0.003\\
$100<\Delta V < 200$ & 0.18$\pm$0.04 & 35$\pm$9 & -0.000$\pm$0.004\\
$200<\Delta V < 300$ & 0.26$\pm$0.08 & 24$\pm$7 & -0.009$\pm$0.004\\
$300<\Delta V < 400$ & 0.12$\pm$0.05 & 37$\pm$18 & -0.018$\pm$0.005\\
$400<\Delta V < 500$ & 0.05$\pm$0.11 & 21$\pm$44 & -0.009$\pm$0.004\\
\hline
\end{tabular}\caption{
Best fit results to the plots of $F(\rp, \rmn{\emph{LWA}})$ in Fig.~\ref{vlim} for a range of $\Delta V$ bins using Eqn.~\ref{fiteq}.
The fits are done using 10 \hkpc bins.}\label{dvfits}
\end{table}

%\subsection{Luminosity dependence of tidal effects}

%-Different changes seen at different separations.

%-Brighter, more massive galaxies, show signs of LWA increase at smaller $\rp$ than fainter galaxies.

%-Luminous galaxies are predominantly found in denser environments than fainter galaxies, thus this result is
%in agreement with previous studies (Lambas 2003,Alonso 2004,2006) who find that tidally induced star formation
%requires smaller $\rp$ in denser environments.

\subsection{Identifying probable interacting galaxies}\label{probpairs}

% This section is based on uncorrected $f(\rp, \rmn{\emph{LWA}})$

As we have seen above, the strongest indication of interacting galaxies in physically associated pairs is an enhanced probability of being classified with \emph{Loose Winding Arms}.  Previously we have studied this signal statistically, averaged over many galaxies in bins of $\rp$. We now attempt to use this morphological signature to identify galaxies which are likely to be interacting, by selecting those which have $f(\rmn{\emph{LWA}})$ above a certain threshold.  After visually examining $\sim100$ galaxies with a range of thresholds we found that $f(\rmn{\emph{LWA}}) > 0.6$ is sufficient to reliably identify galaxies with tidal features, provided there are at least two \emph{LWA} votes. (A single vote may occasionally be spurious.)  Given the roughly similar number of times each object has been classified, the requirement of at least two \emph{LWA} votes may be expressed as a threshold on $p(\bmath{M})$.  In this case,  $p(\bmath{M}) = p(\rmn{\emph{Features}} \cap \rmn{\emph{Edge-on
 =No}} \cap \rmn{\emph{Spiral=Yes}}) = f(\rmn{\emph{Features}}) f(\rmn{\emph{Edge-on=No}}) f(\rmn{\emph{Spiral=Yes}})$, which we hereafter refer to as $p(\rmn{\emph{FNS}})$.  Most GZ2 objects have at least 30 classifications, so applying a threshold of $p(\rmn{\emph{FNS}}) > 0.1$ means that the question \emph{How tight are the spiral arms?} has received at least three answers, at least two of which must have indicated \emph{Loose Winding Arms} if $f(\rmn{\emph{LWA}}) > 0.6$.  As described in Section \ref{morphprob}, we denote this selection as $f(\rmn{\emph{LWA}}\,|\, p(\rmn{\emph{FNS}}) > 0.1) > 0.6$.

The chosen thresholds are a reasonable compromise.  If we choose lower thresholds we select potential interacting galaxies with more subtle tidal features, but at the same time increase the contamination from non-interacting pairs.  The number of galaxies which satisfy our criteria will inevitably represent only a fraction of the galaxies which are truly interacting.  Some galaxies may possess tidal features that are unobservable (i.e. do not result in a significantly elevated $f(\rmn{\emph{LWA}}$) due to their low surface brightness, an unfavourable sky orientation (i.e. edge-on galaxies), or other observational limitation.  Other truly interacting galaxies may possess only weak, or entirely absent, morphological signatures of that interaction, due to the orbital parameters of the interaction. As discussed by \citet{toom1972}, galaxies which are rotating in the same sense as the orbital pass of the companion will form tidal arms which are much more pronounced than if the galaxy is rotating in the opposite direction.  Finally, galaxies which display \emph{Loose Winding Arms} have probably already undergone at least one close pass of their companion.  Some close pairs will be in the early stages of their interaction and thus have not yet formed tidal arms.

We visually examined all of the galaxies in pairs with $\rp < 200$ \hkpc and $\Delta V < 500$ \kms and selected to be interacting with $f(\rmn{\emph{LWA}}\,|\, p(\rmn{\emph{FNS}}) > 0.1) > 0.6$.  Essentially all of these objects show obvious signs that they are interacting with a companion.  In many cases the companion is also selected as interacting.
However, galaxies can belong to multiple pairs and most of the interacting galaxies in large separation ($\rp \ga 100$ \hkpc) pairs were
actually found to have a closer companion that is likely to be the true cause of the interaction.  In these cases the companion at larger separation typically shows no signs of interaction, and may not be physically associated. This is also the case for some very close pairs, where the closest companion is actually an interloper, and the interacting companion is at a larger projected separation. Also, due to the redshift incompleteness of the SDSS, especially at small angular separations, some of the interacting companions of galaxies identified as having \emph{Loose Winding Arms} will be missing from our sample.

Table \ref{lwaimages} shows examples of galaxies with \emph{Loose Winding Arms}, i.e. $f(\rmn{\emph{LWA}}\,|\, p(\rmn{\emph{FNS}}) > 0.1) > 0.6$, for a range of $\rp$ up to $100$ \hkpc.  We see that the high mass pairs consist mostly of early type galaxies, while the lower mass pairs are mostly mixed and late-type pairs with bluer colours.

%-This method allows us to select interacting galaxies up to large separation,.. up to the maximum separation between 1st and 2nd pass...

\begin{table*}
\setlength{\tabcolsep}{0pt}
\begin{tabular}{c c c c c c c c c c c}
\hline $\rp$ (\hkpc) & 5 & 15 & 25 & 35 & 45 & 55 & 65 & 75 & 85 & 95 \\

\hline $M_*$ ($M_{\sun}$) & 11.0--10.4 & 10.9--10.8 & 10.9--11.1 & 10.7--11.0 & 10.7--11.1 & 10.8--10.9 & 11.0--11.2 & 11.0--11.1 & 10.8--11.1 & 10.9--11.2\\
&
\includegraphics[height=0.64in]{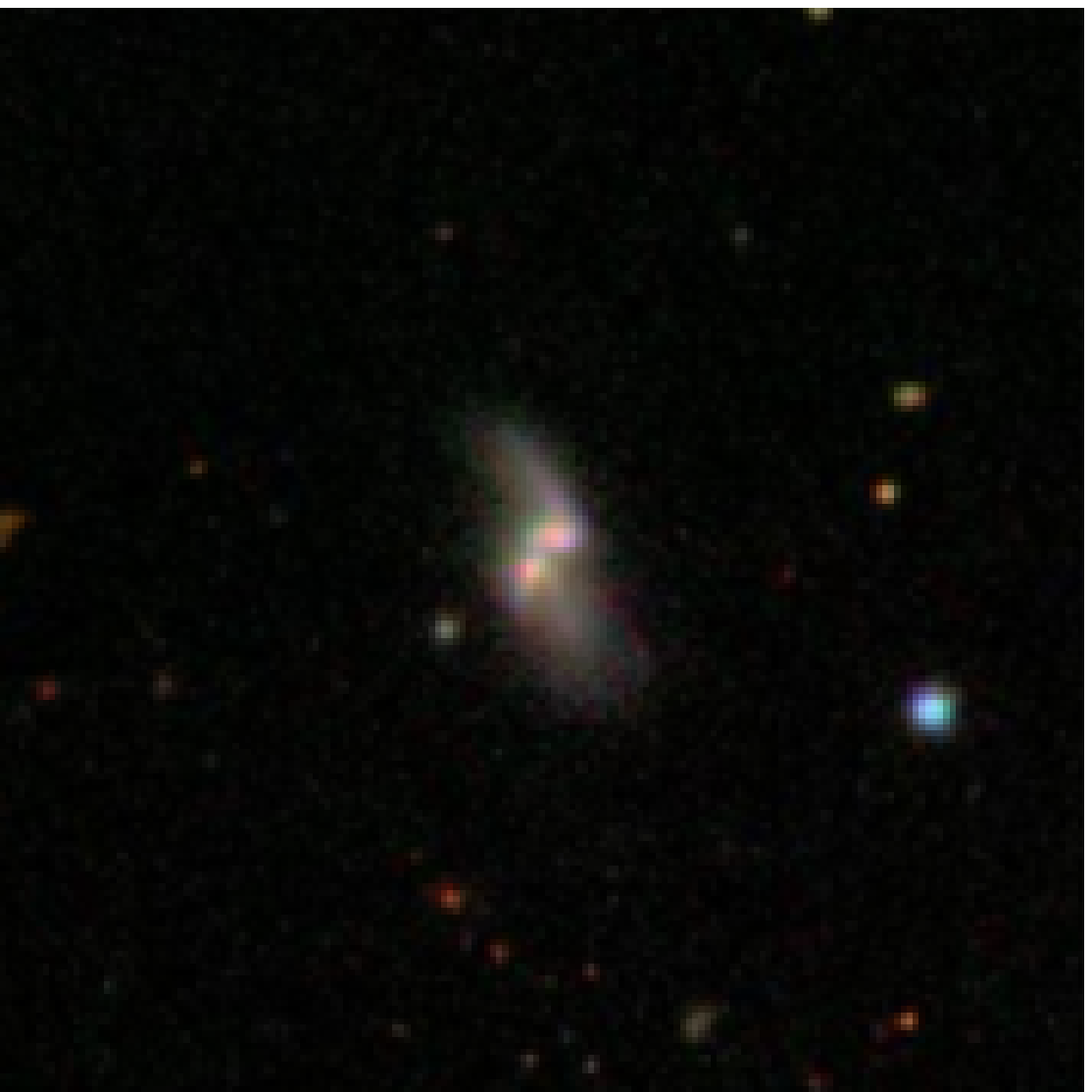}&
\includegraphics[height=0.64in]{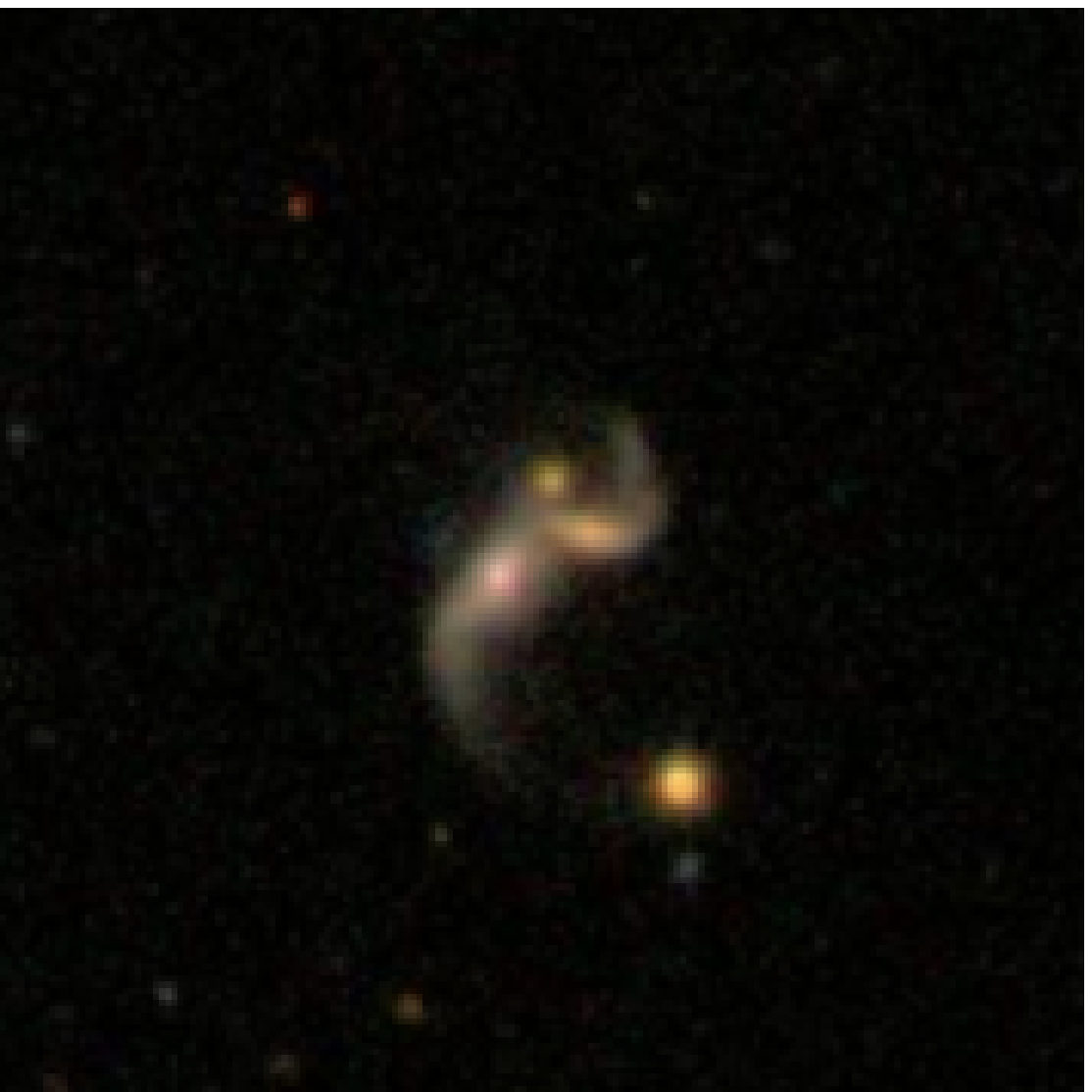}&
\includegraphics[height=0.64in]{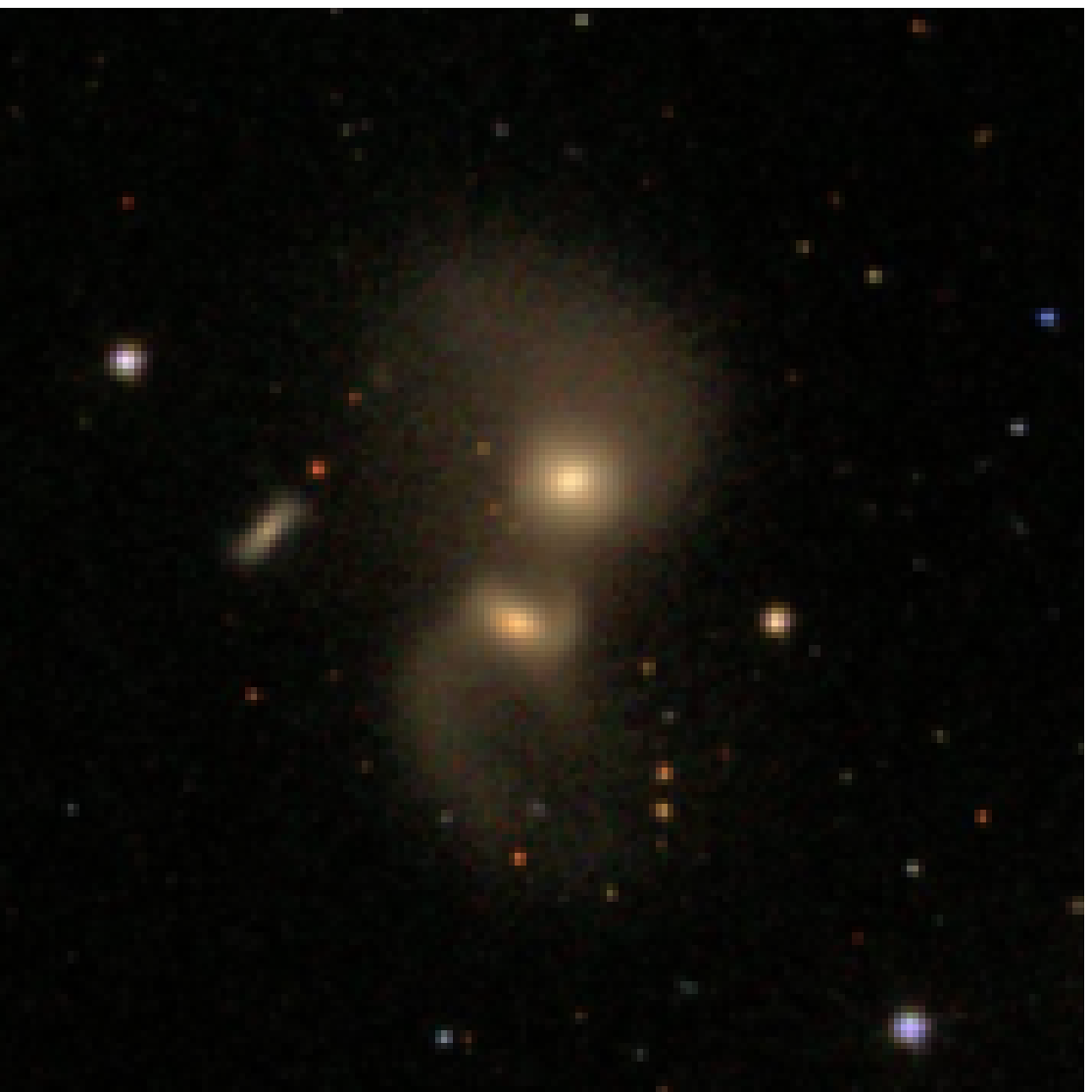}&
\includegraphics[height=0.64in]{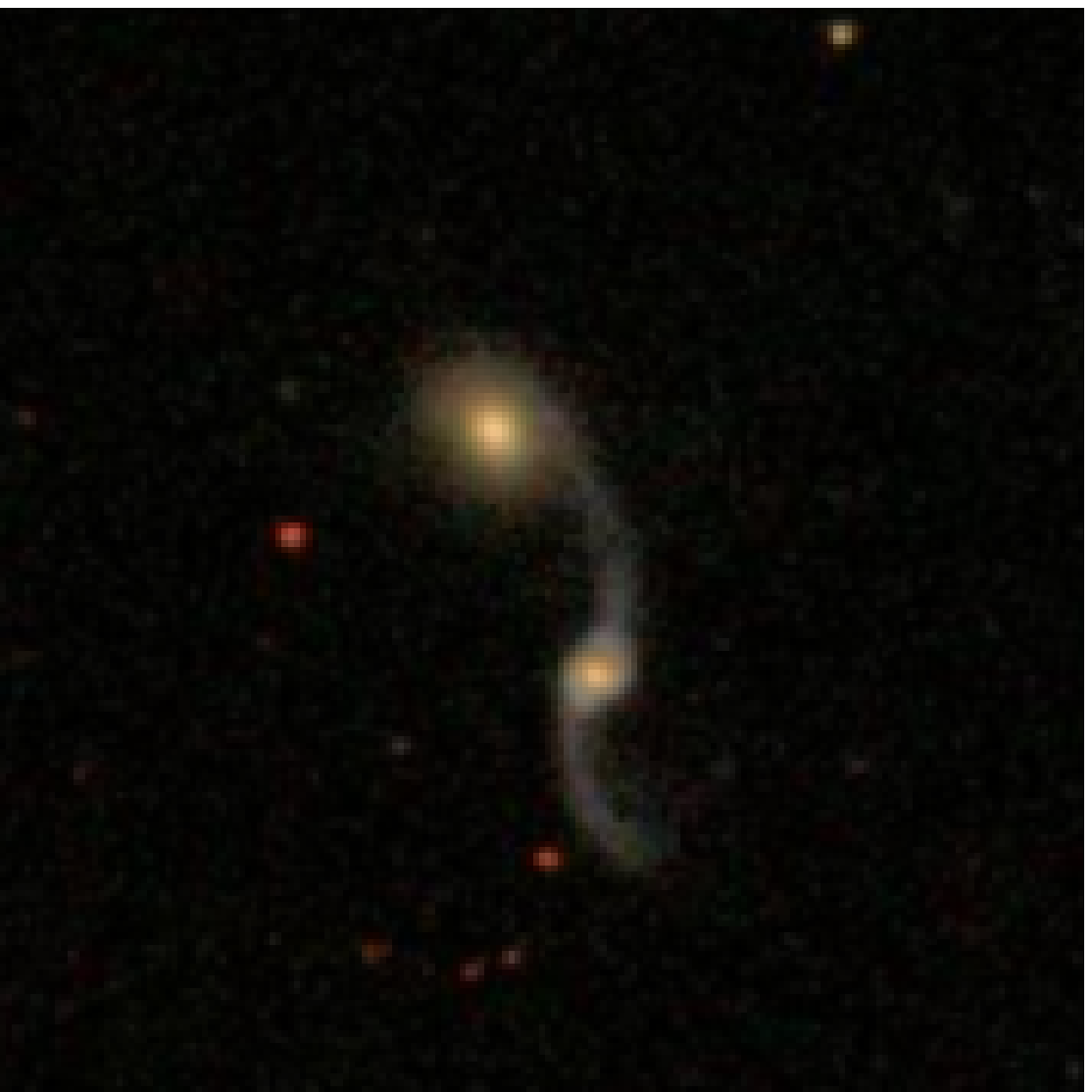}&
\includegraphics[height=0.64in]{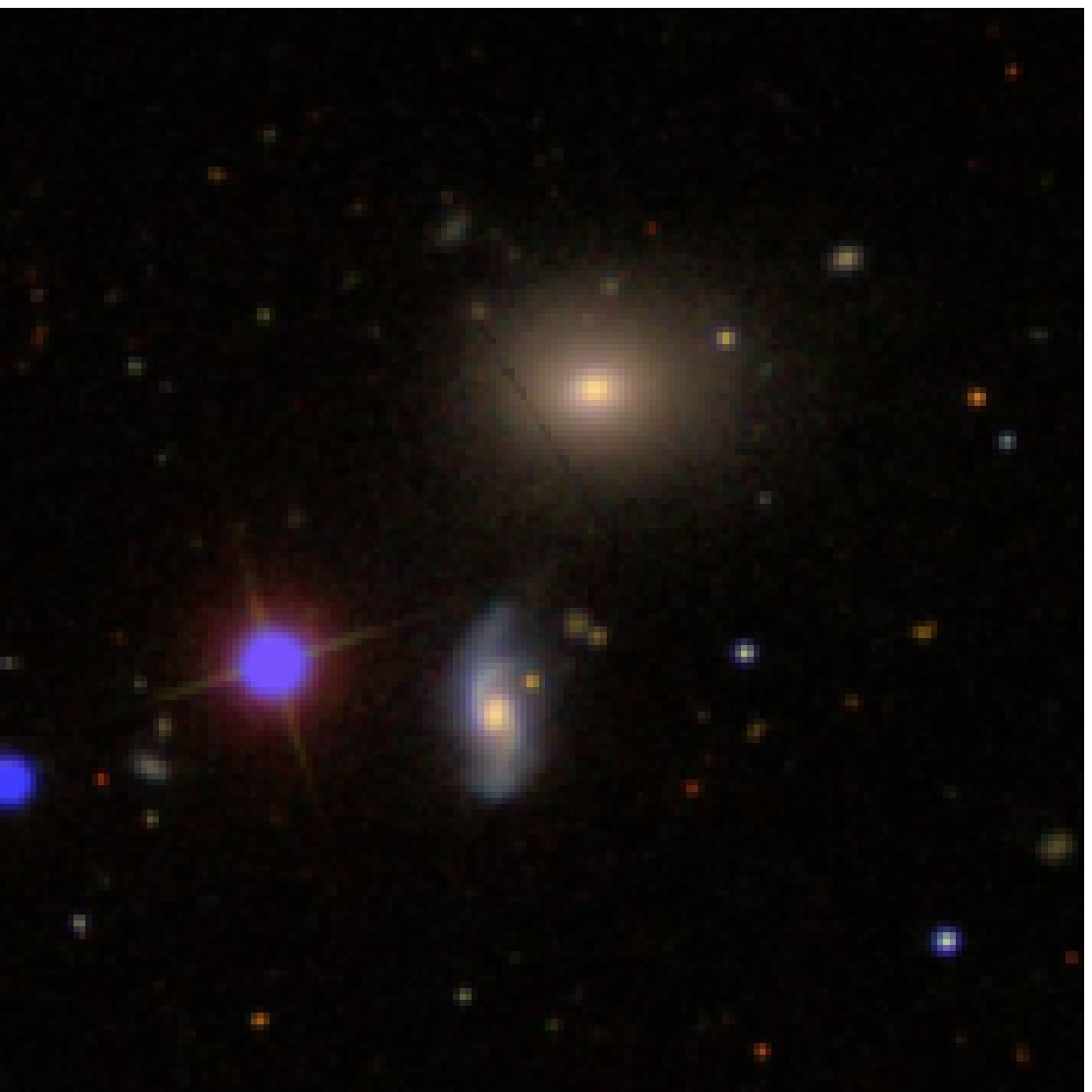}&
\includegraphics[height=0.64in]{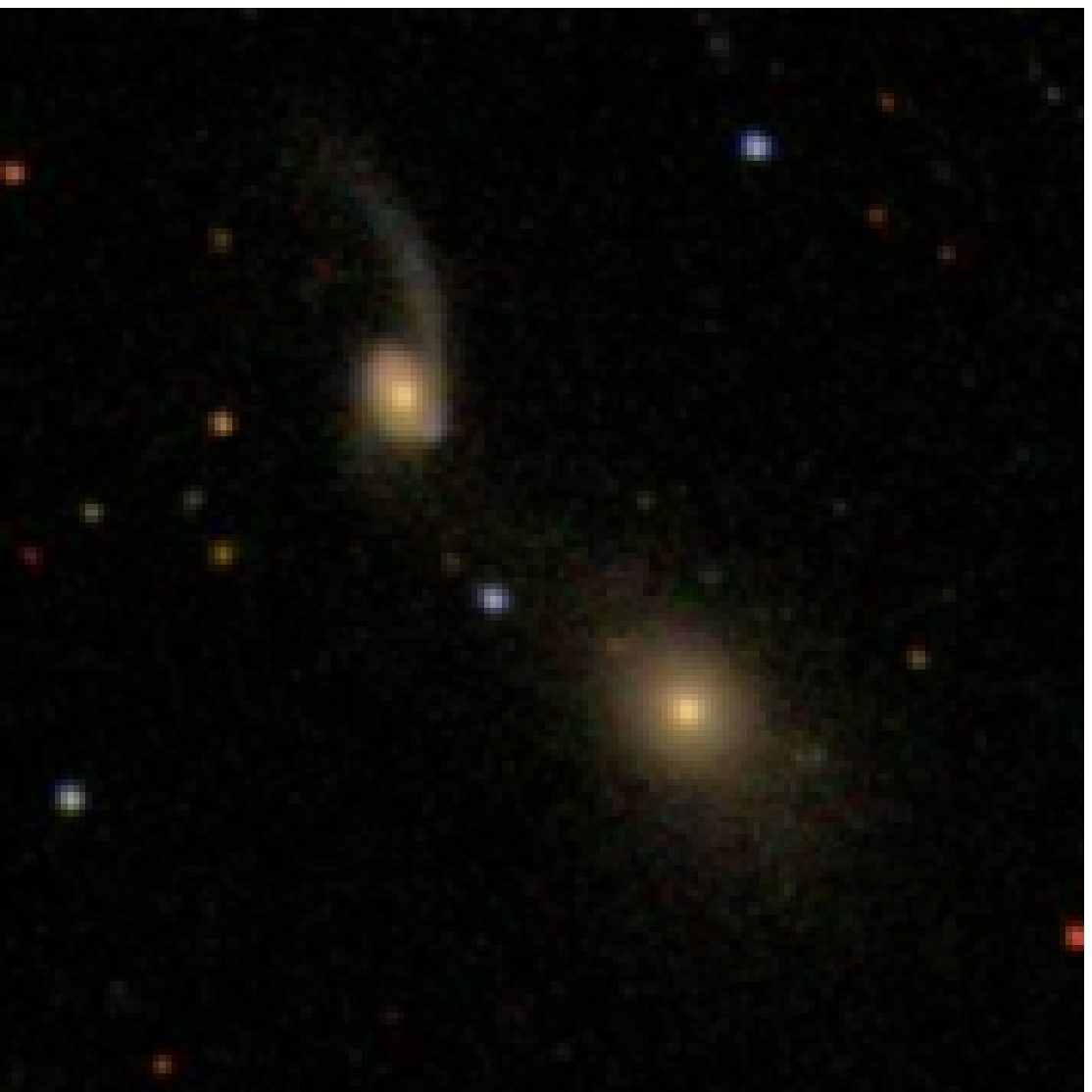}&
\includegraphics[height=0.64in]{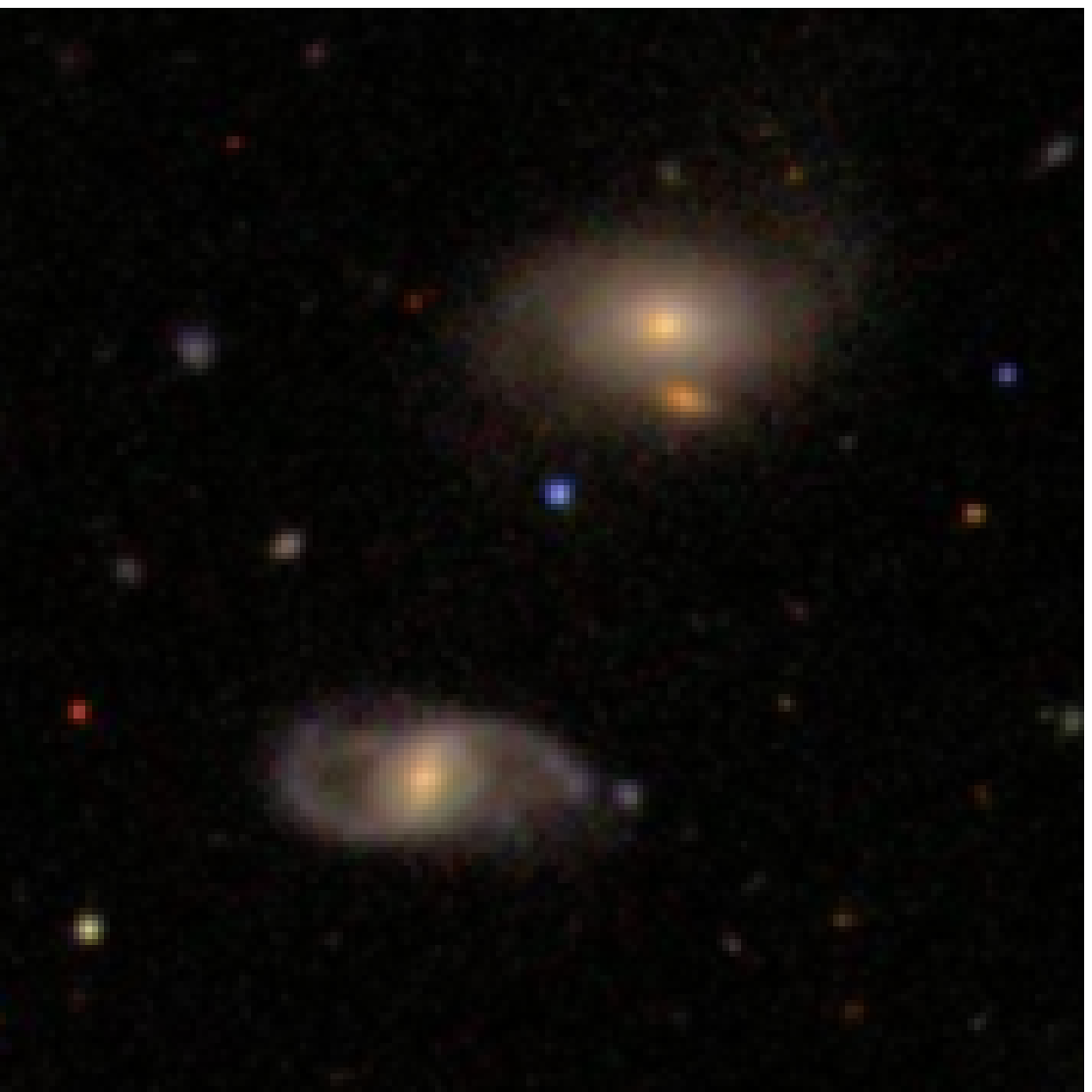}&
\includegraphics[height=0.64in]{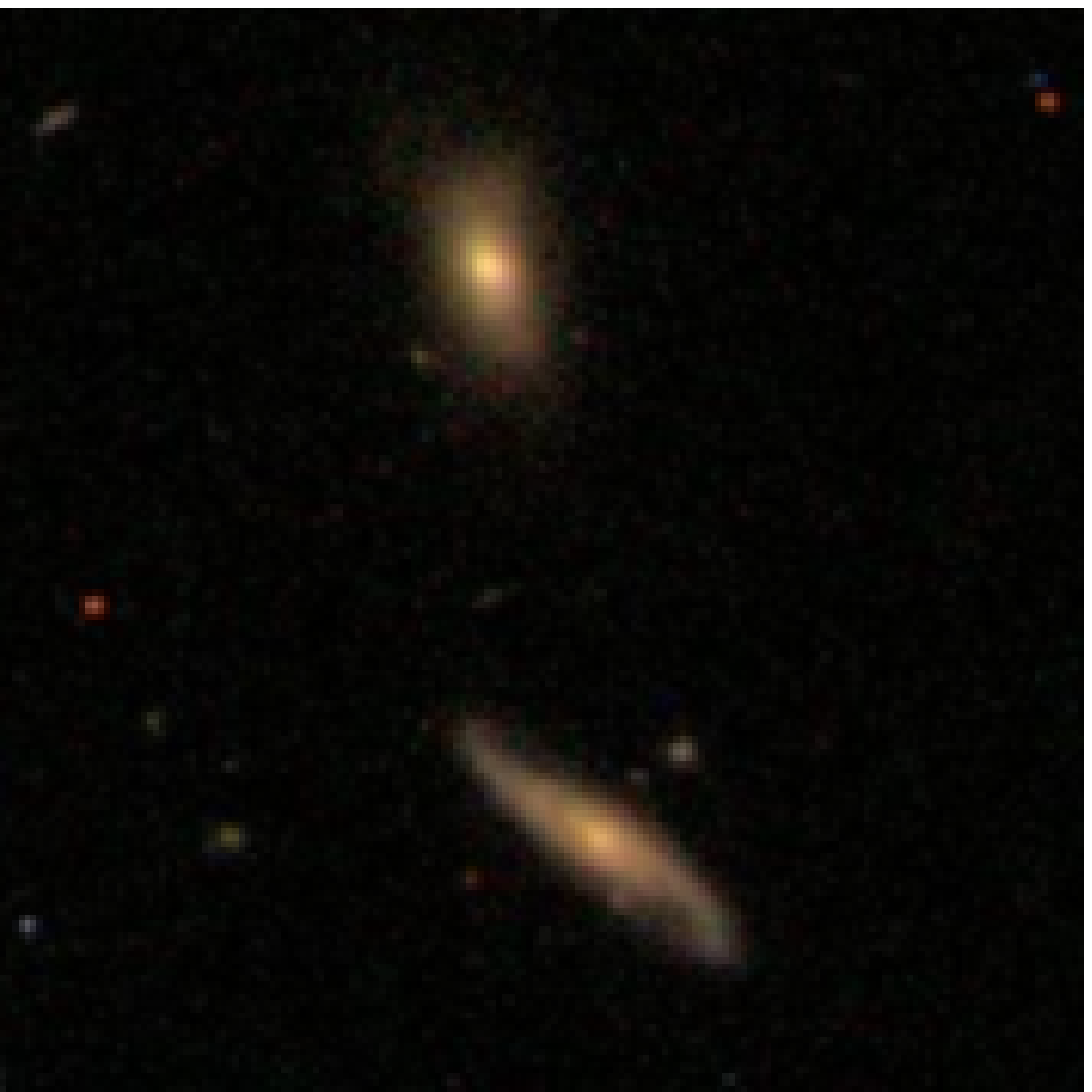}&
\includegraphics[height=0.64in]{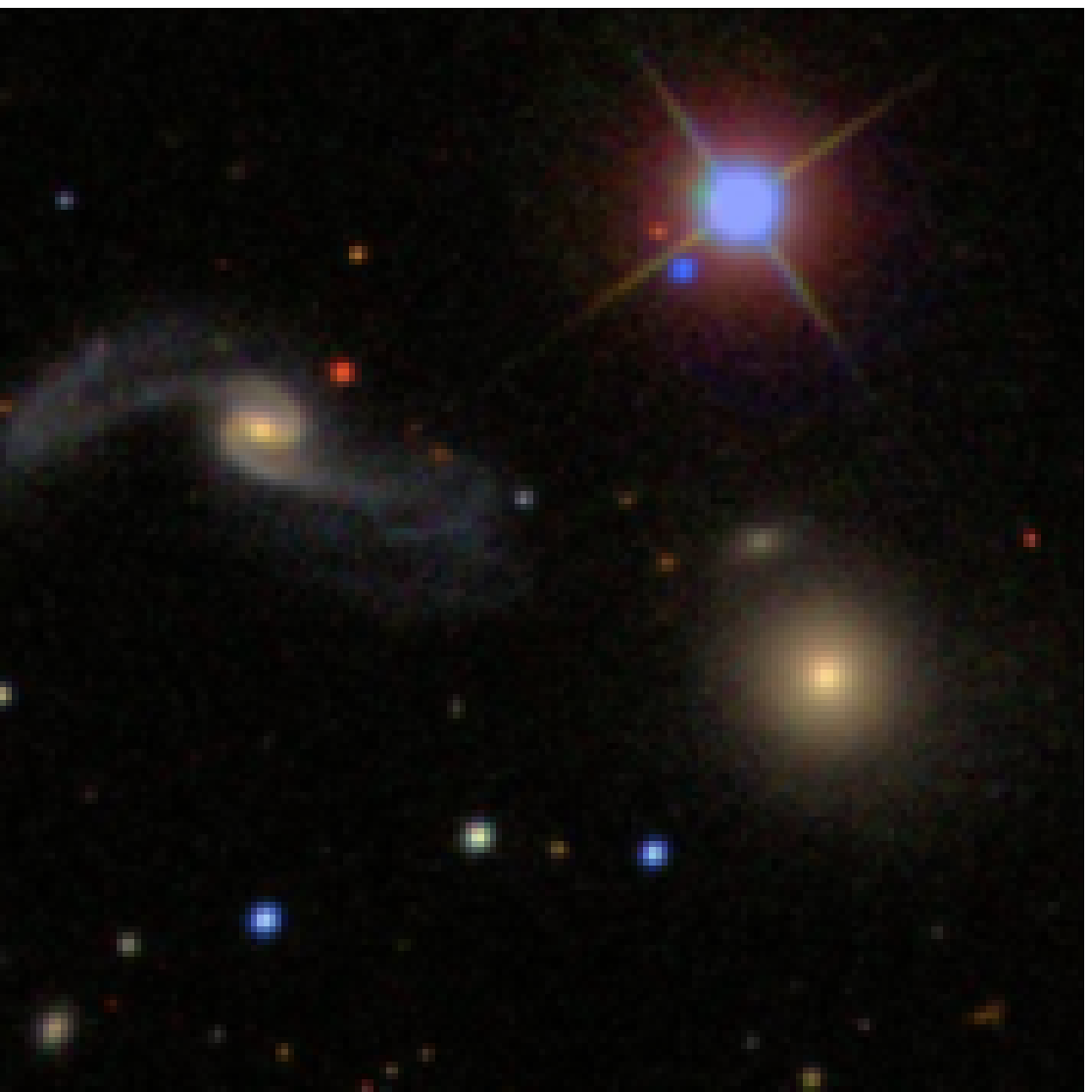}&
\includegraphics[height=0.64in]{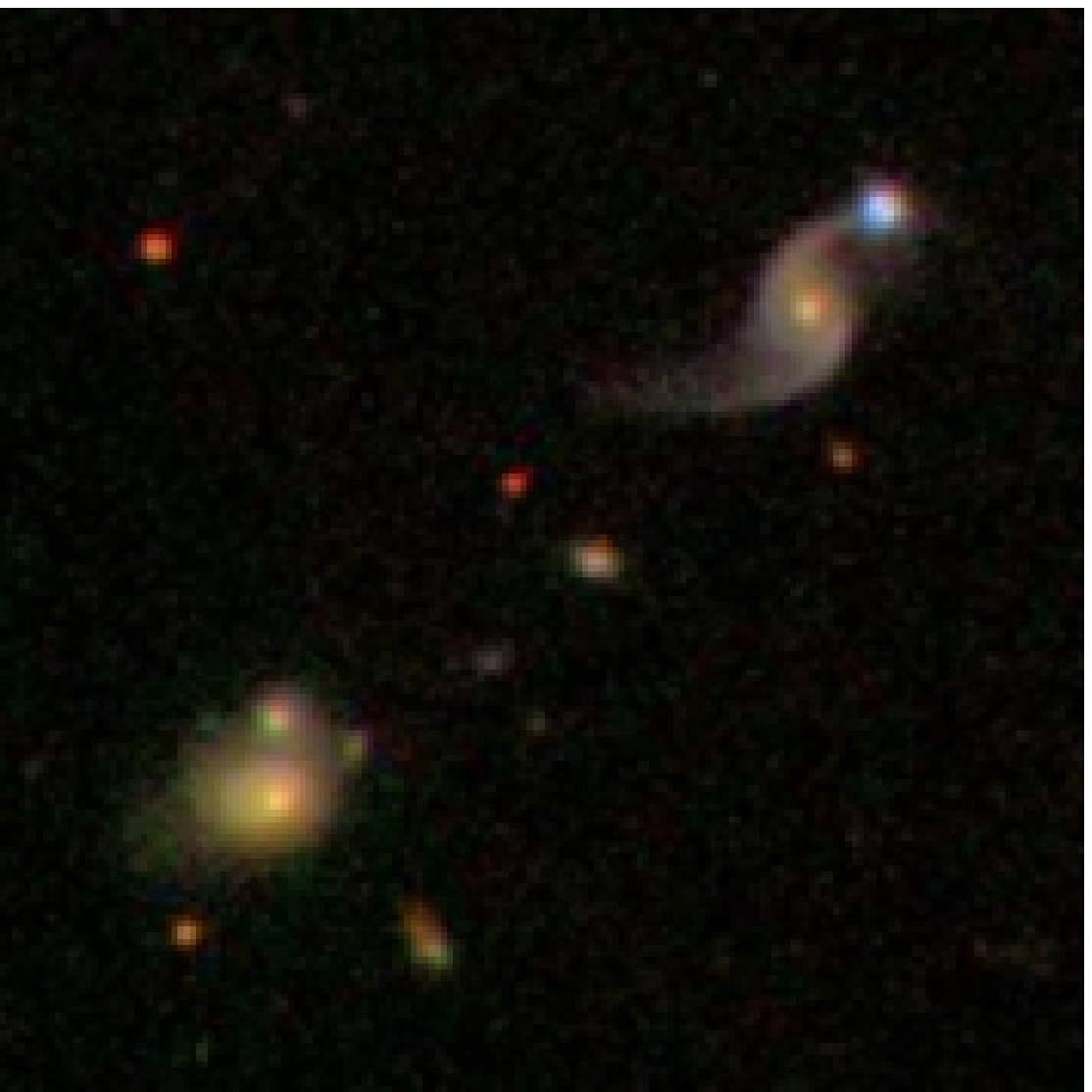}\\

\hline $M_*$ ($M_{\sun}$)& 9.9--10.1 & 9.5--9.5 & 10.0--10.4 & 10.0--10.3 & 10.0--10.1 & 9.8--9.6 & 9.6--9.9 & 9.8--10.1 & 9.7--9.8 & 9.8--10.2\\
&
\includegraphics[height=0.64in]{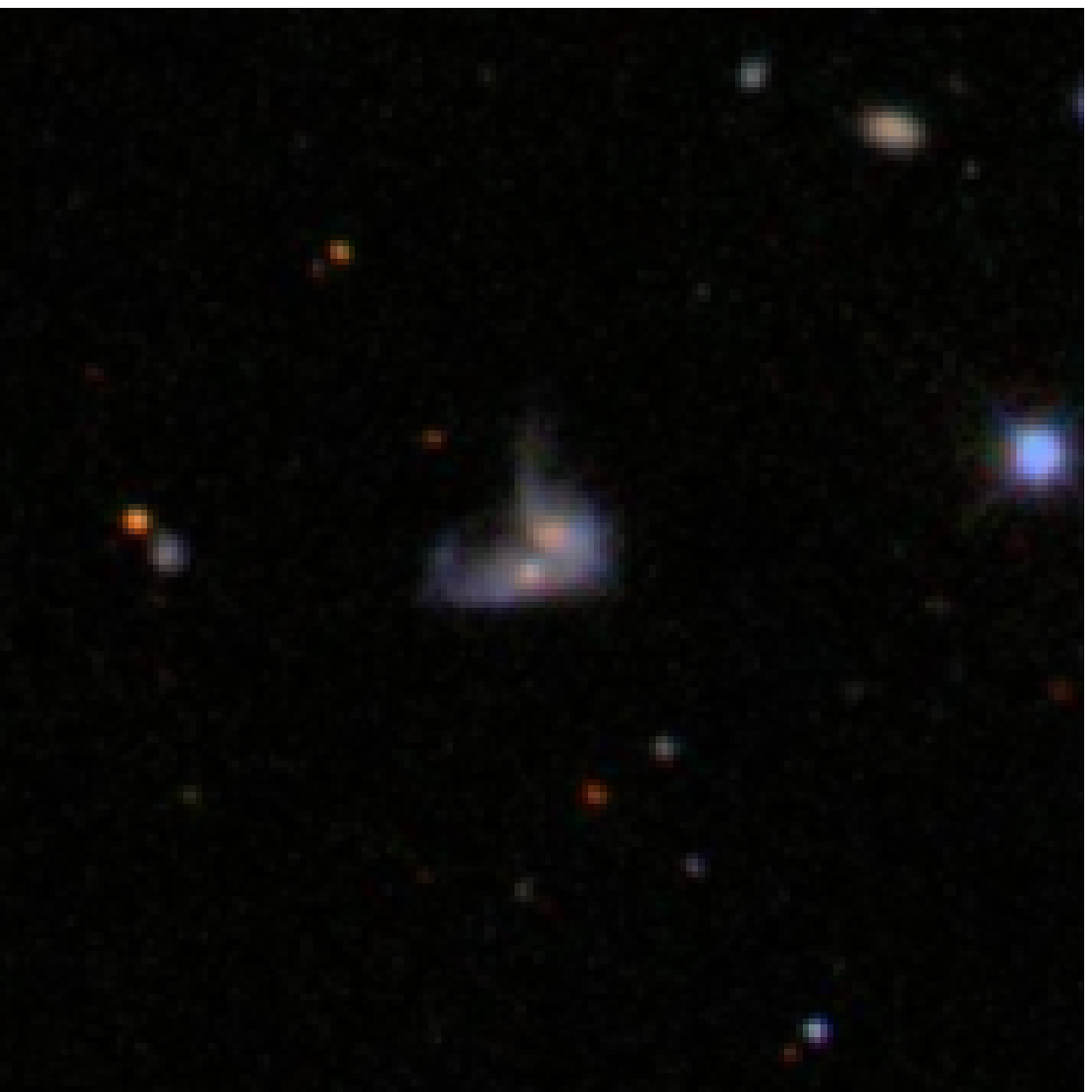}&
\includegraphics[height=0.64in]{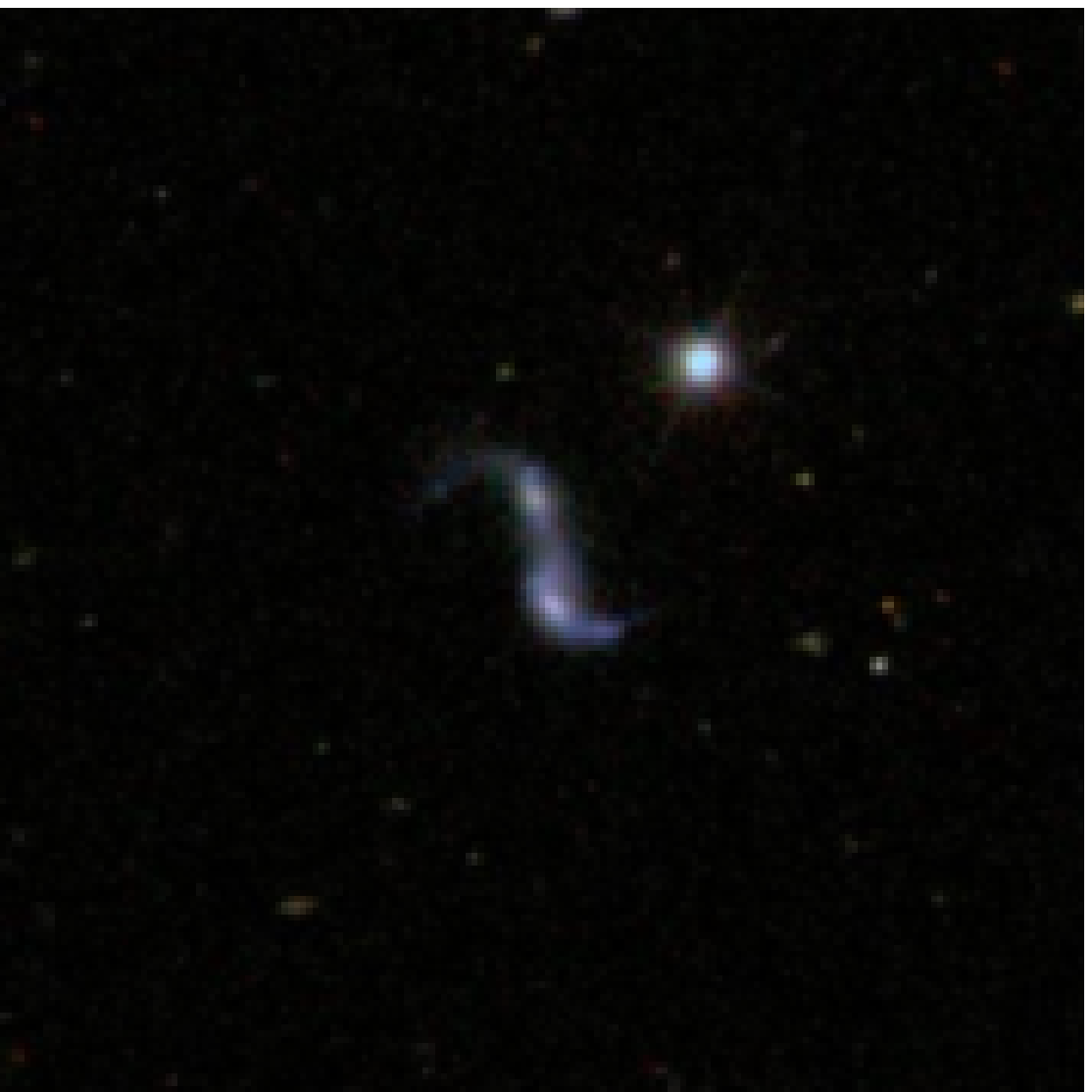}&
\includegraphics[height=0.64in]{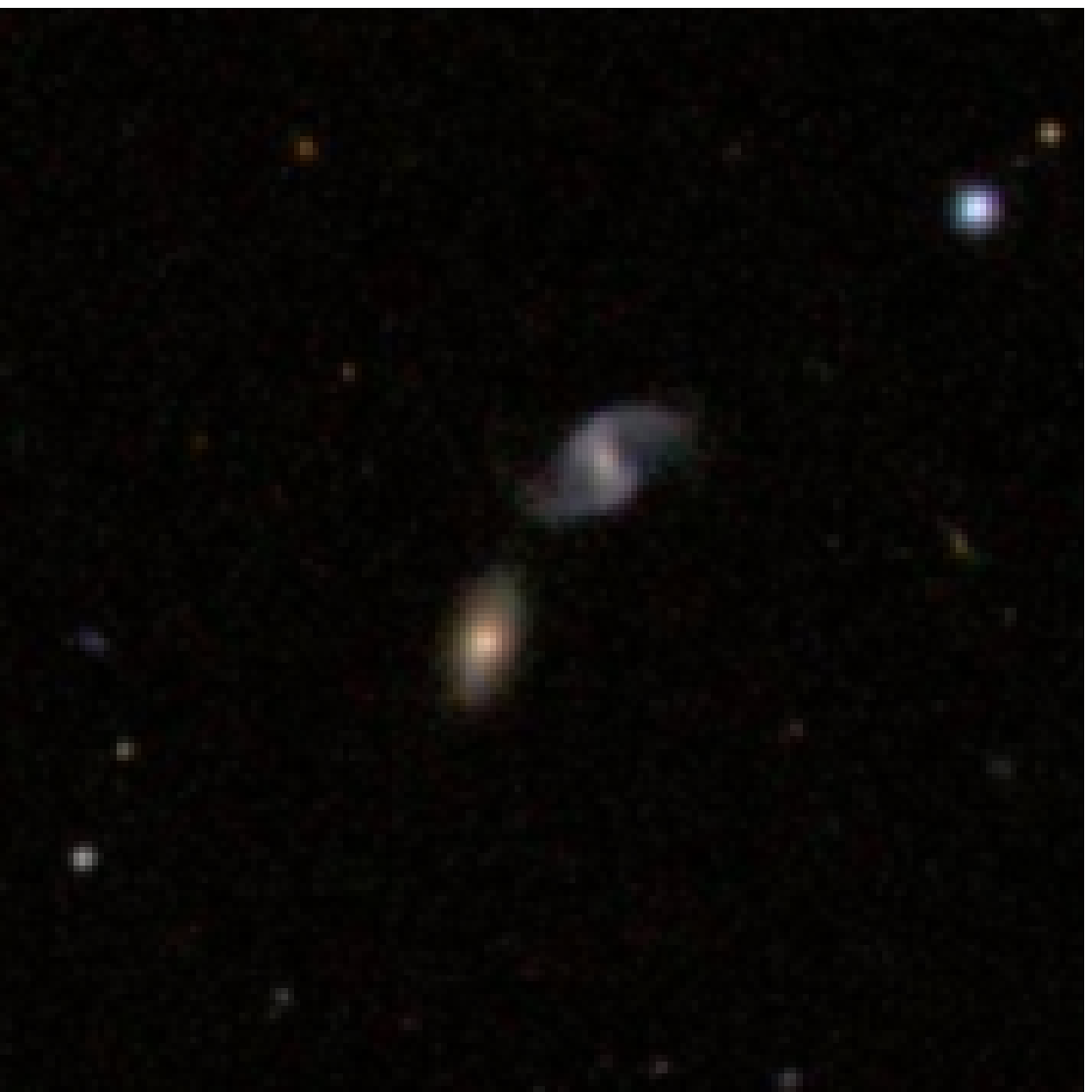}&
\includegraphics[height=0.64in]{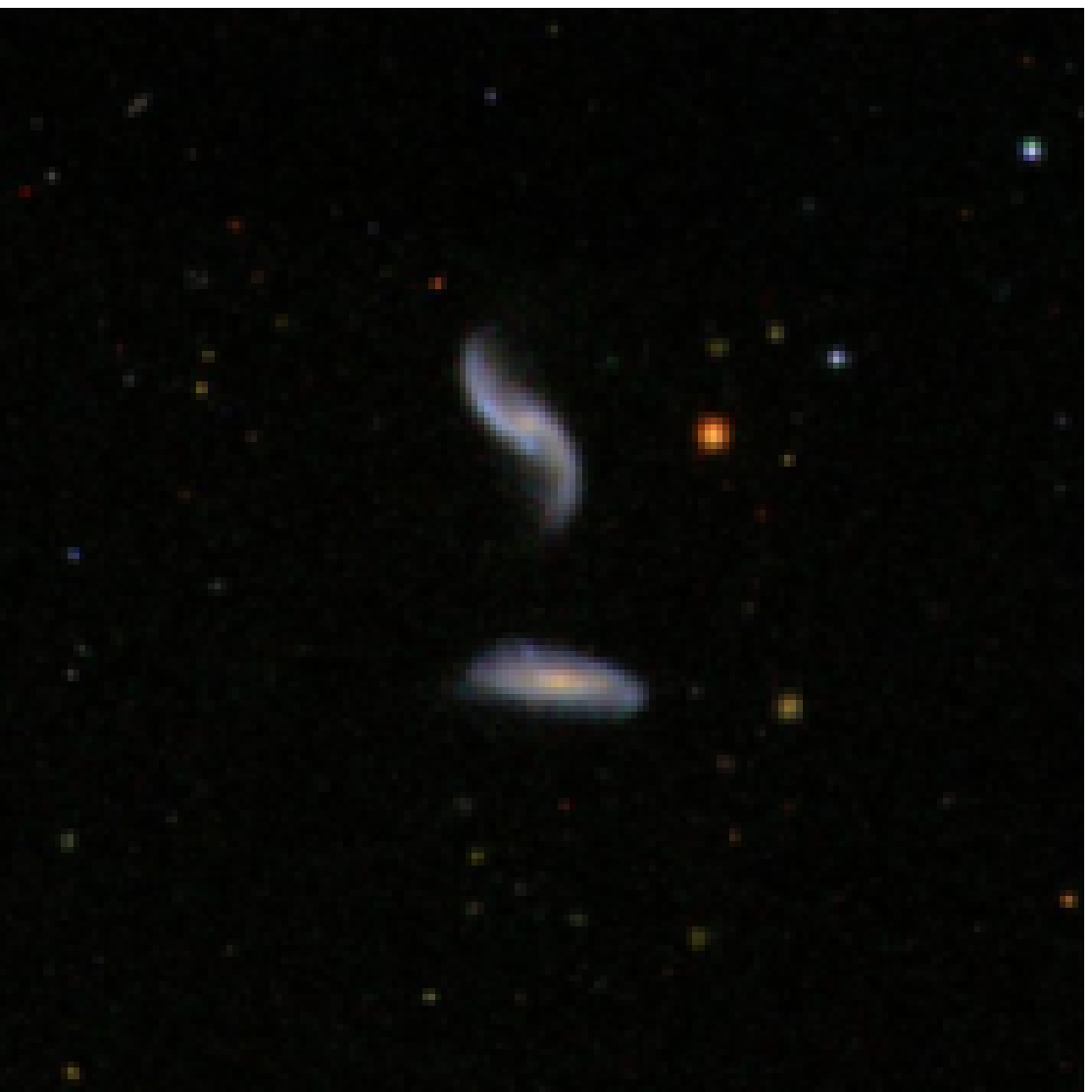}&
\includegraphics[height=0.64in]{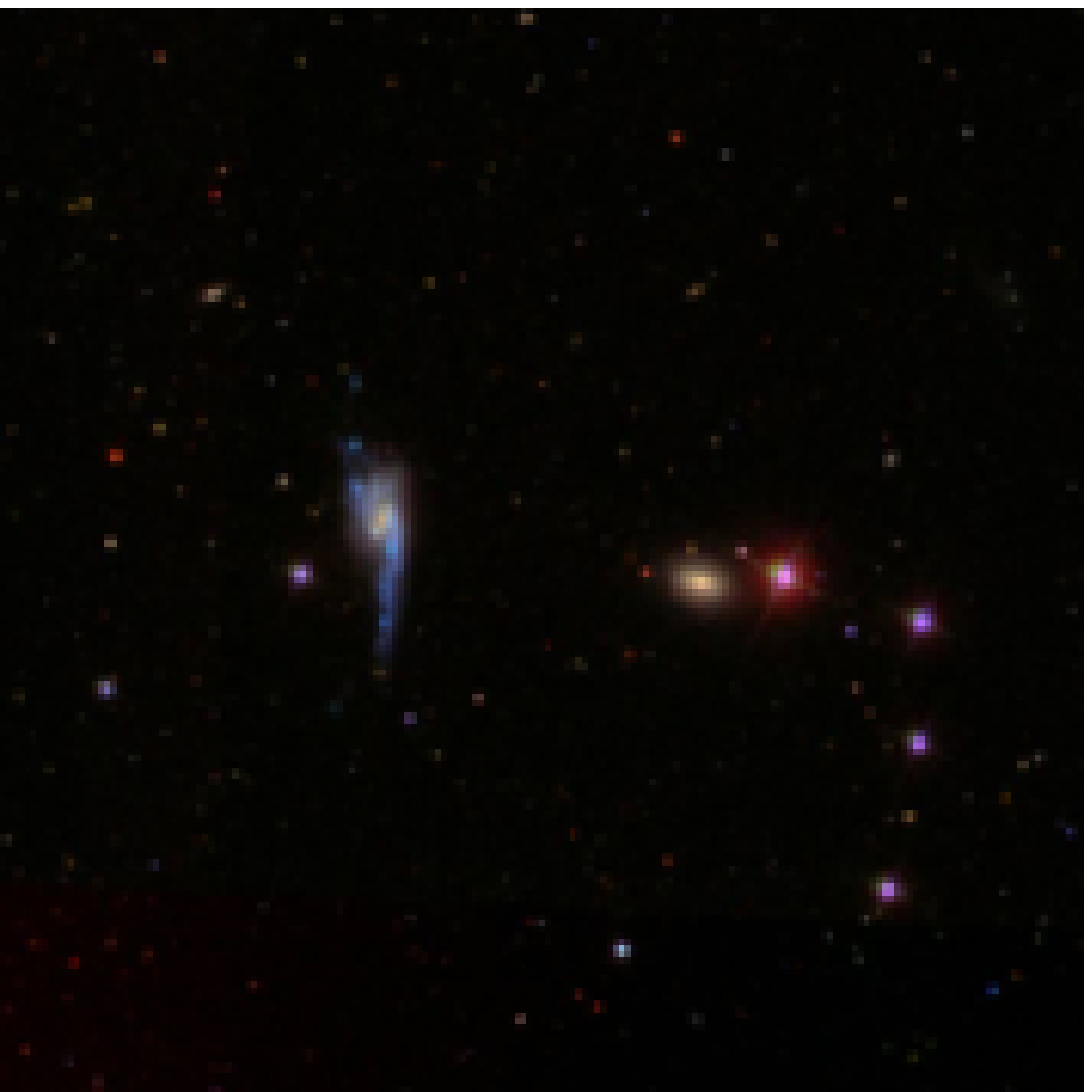}&
\includegraphics[height=0.64in]{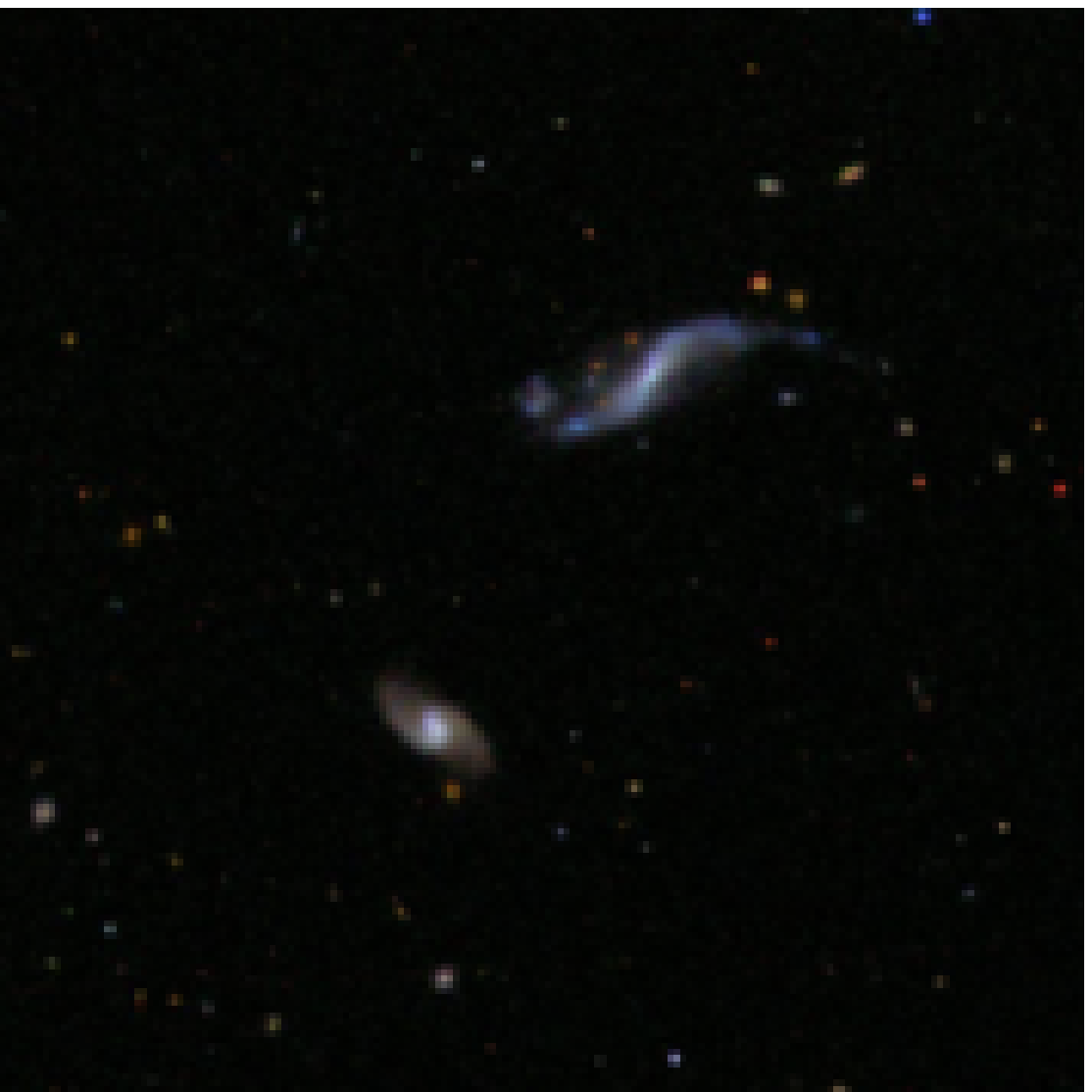}&
\includegraphics[height=0.64in]{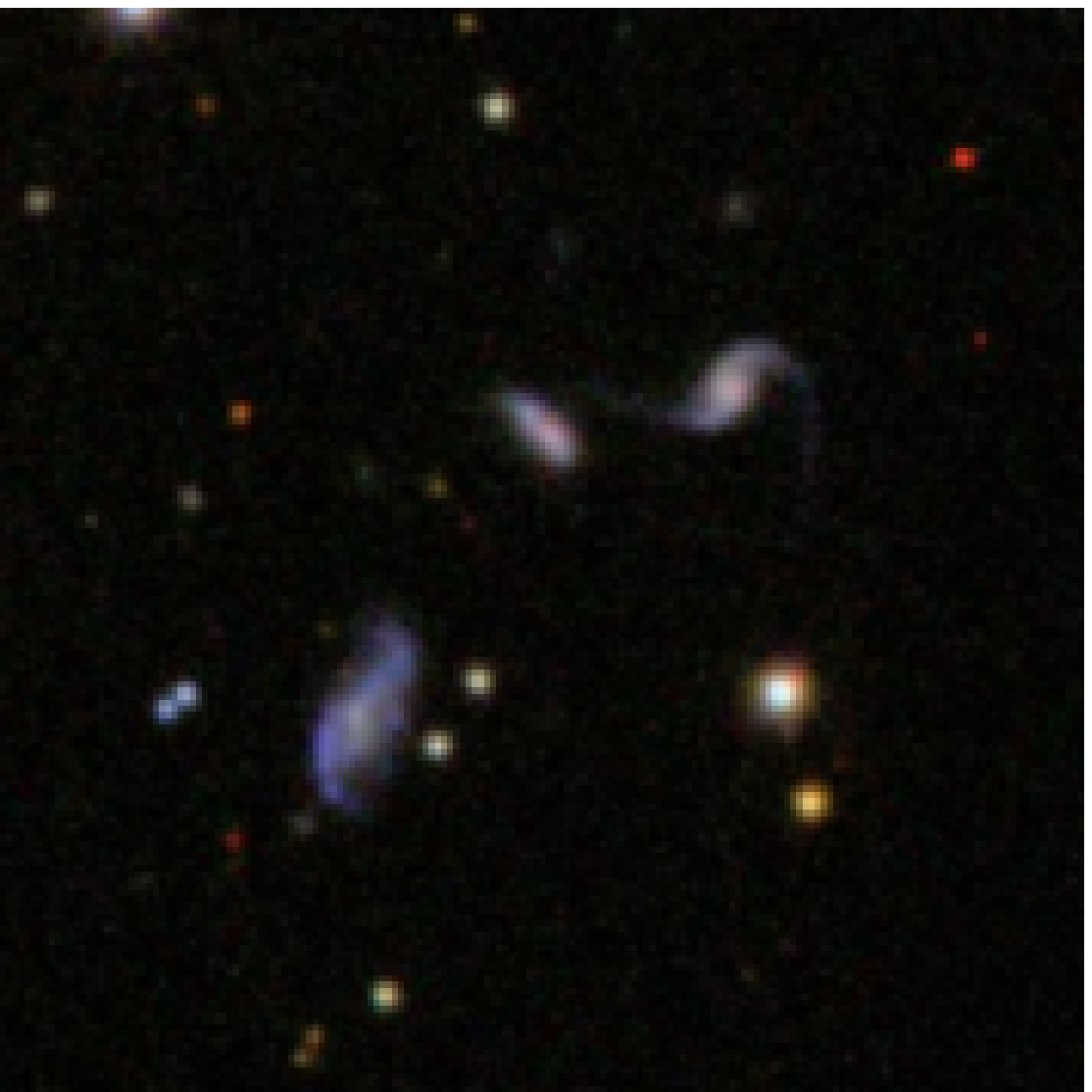}&
\includegraphics[height=0.64in]{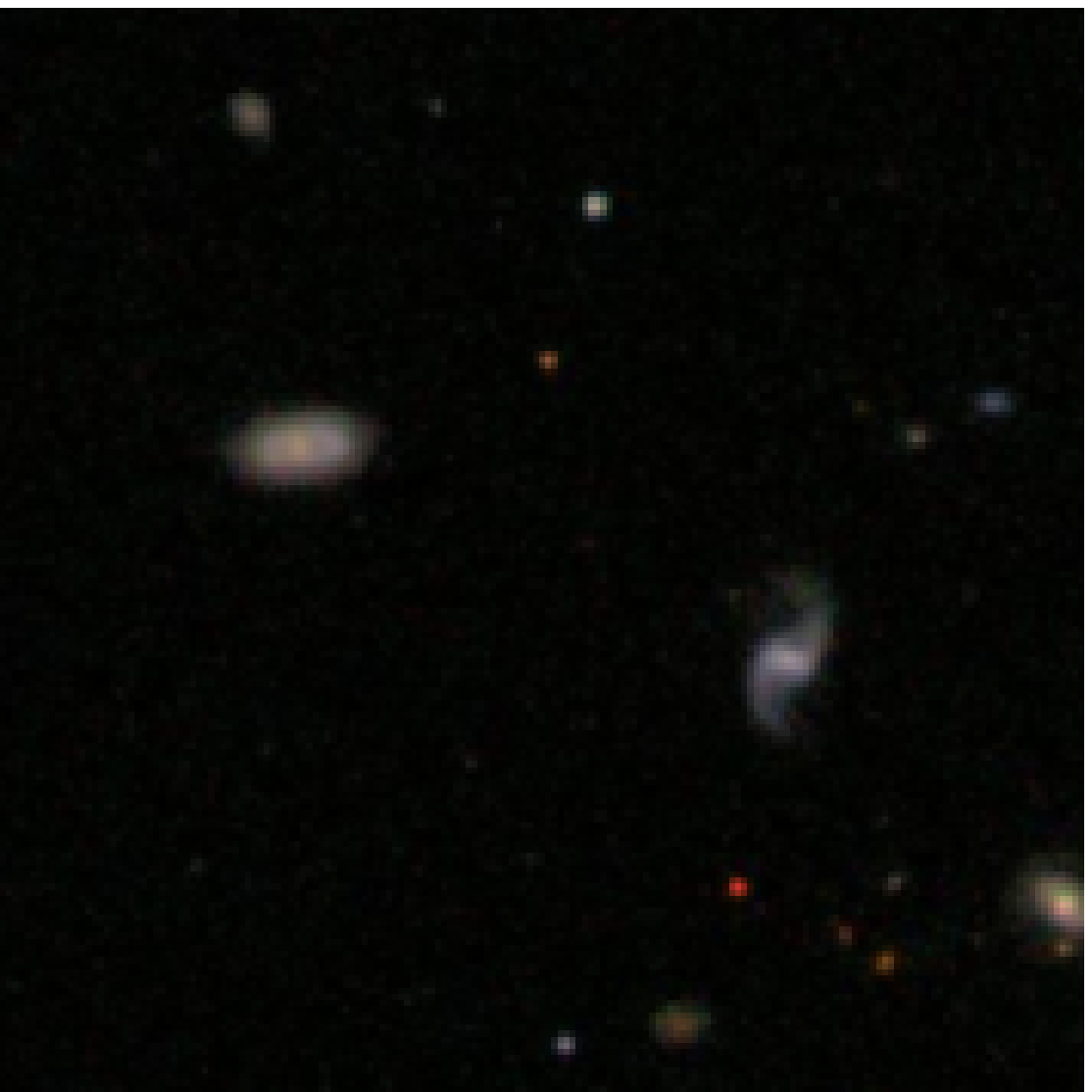}&
\includegraphics[height=0.64in]{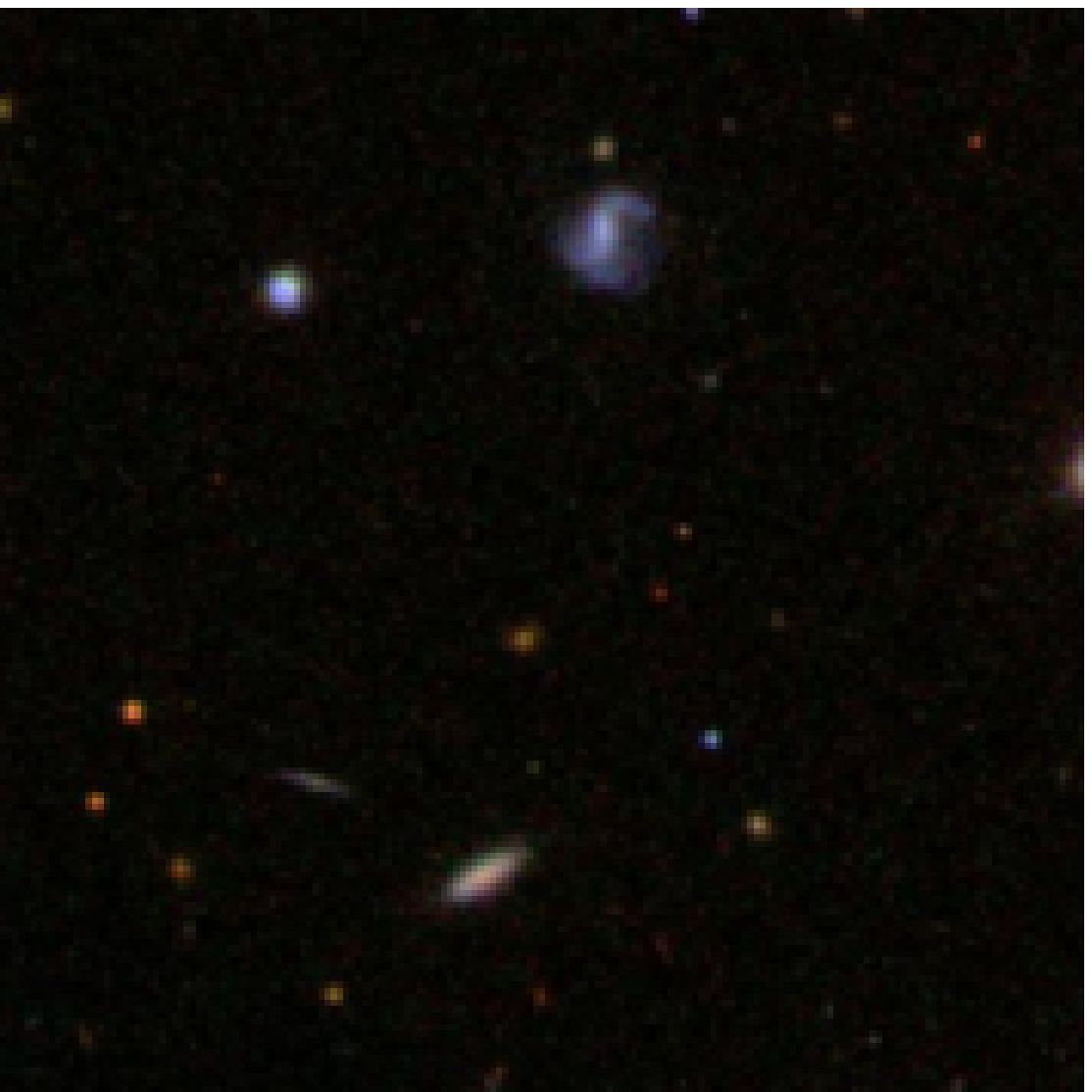}&
\includegraphics[height=0.64in]{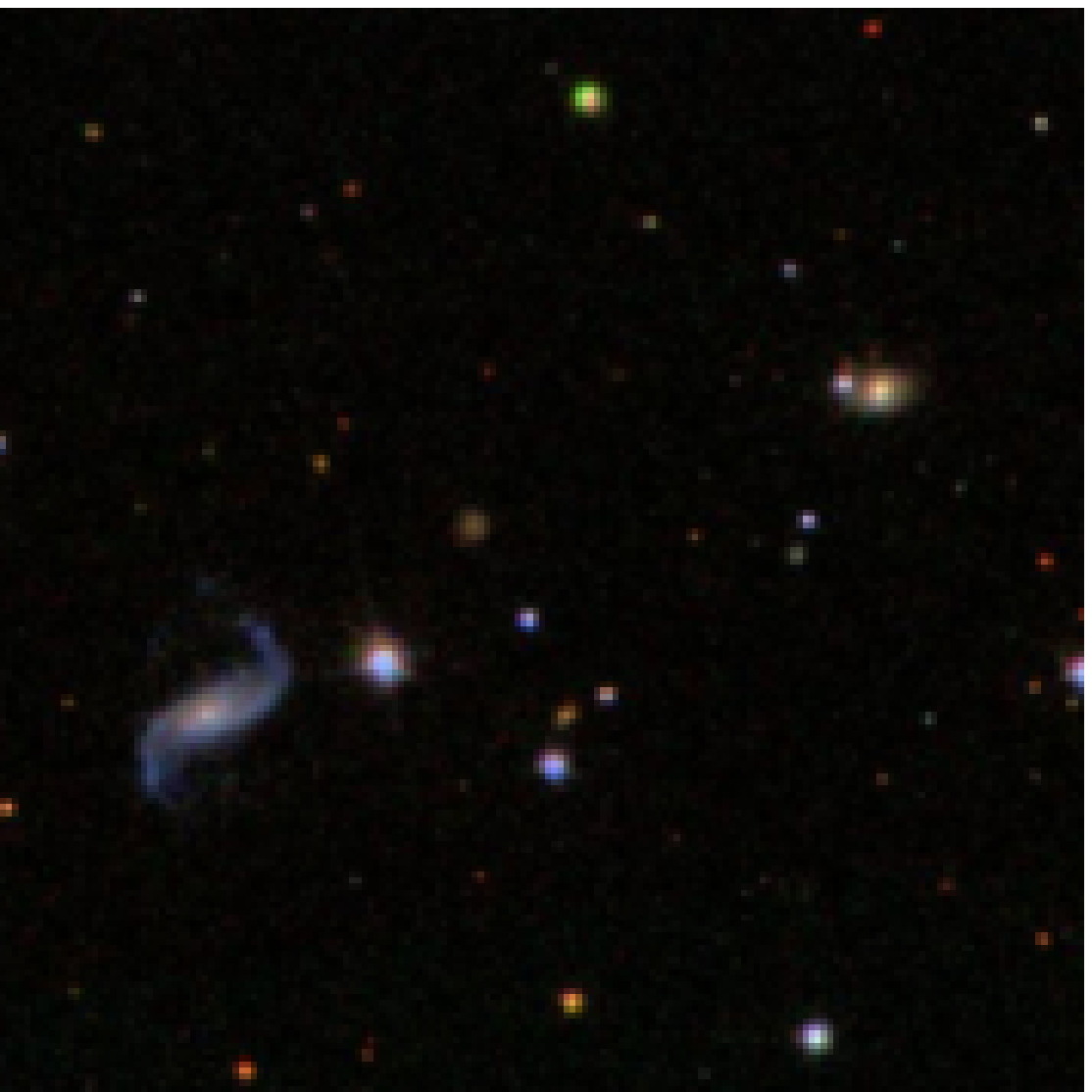}\\

\hline

\end{tabular}
  \caption{Galaxies selected to be interacting members of a pair, using $f(\rmn{\emph{LWA}}\,|\, p(\rmn{\emph{FNS}}) > 0.1) > 0.6$ and $\Delta V < 500$ \kms. The figures along the top indicate the approximate $\rp$ (in \hkpc) of the pairs in each column.  The other figures give the stellar masses (in $M_{\sun}$) of galaxies in the adjacent image.  The top and bottom rows present examples with high and low stellar mass, respectively.  The width of these figures is $\sim 110$ \hkpc, based on the average redshift of the pair.
}\label{lwaimages}
\end{table*}

\subsection{Accounting for interlopers in the frequency of galaxy pair interactions}\label{interlopers}

In this section we demonstrate how morphological indicators of interaction may be used to to refine estimates of the frequency of galaxy pair interactions.  We employ a mass-limited sample, selected by imposing a minimum mass limit of $10^{9.5} M_{\sun}$ and redshift limits of $0.01<z<0.05$, giving a sample of $44064$ galaxies.  The highest mass galaxies in our sample have $M_* \sim 10^{11.5} M_{\sun}$.  We consider all galaxy pairs with $\Delta V < 500$ \kms and $\rp < 300$ \hkpc, in $5$ \hkpc bins.

As we have seen, a significant fraction of the galaxies in these pairs will not be truly interacting.
Previous studies have found this contamination to a strong function of projected separation \citep{alon2004,pere2006} and galaxy luminosity (and thus mass), with the faintest pairs being most affected \citep{patt2008}.
We therefore use the morphological indicators of interactions, explored above, to estimate and correct for the influence of interlopers on our determination of the interaction rate.

We calculate the average number of companions at separation $\rp$ (and within $\Delta V < 500$ \kms) for each galaxy in the sample, $\Nc(\rp)$, following the method in Section \ref{nccalc} (with $w_\rmn{int} = 1$).  This method accounts for the various sample selection issues to produce a corrected estimate of the average number of close companions per galaxy.  As some of these companions will not be physically interacting, this is an upper limit on the number of interacting companions per galaxy.

We also measure the average number of companions in pairs displaying \emph{Loose Winding Arms}, ${\Nc}_\rmn{\emph{{LWA}}}(\rp)$. This is achieved in an identical fashion, but limited to probable interacting galaxies where at least one member of a pair is identified as having \emph{Loose Winding Arms}, using the criteria developed in Section \ref{probpairs}, i.e. $w_\rmn{int} = 1$ if $f(\rmn{\emph{LWA}}\,|\, p(\rmn{\emph{FNS}}) > 0.1) > 0.6$ for either member of the pair and $w_\rmn{int} = 0$ otherwise.

The fraction of companion galaxies which show indications of possible interaction (i.e. with $f(\rmn{\emph{LWA}}\,|\, p(\rmn{\emph{FNS}}) > 0.1) > 0.6$), can be calculated for each $\rp$ bin using,
\begin{equation}\label{interlopereq}	
  \mathcal{F}_\rmn{\emph{{LWA}}}(\rp) = \frac{{\Nc}_\rmn{\emph{{LWA}}}(\rp)}{\Nc(\rp)}\,.
\end{equation}
This fraction is related to the true interacting fraction, $\mathcal{F}_\rmn{int}(\rp)$, such that
\begin{equation}\label{finteq}
  \mathcal{F}_\rmn{int}(\rp) = \frac{\mathcal{F}_\rmn{\emph{{LWA}}}(\rp) - \mathcal{F}_\rmn{\emph{{LWA}}}(\rp \to \infty)}{P_\rmn{int obs}} \,,
\end{equation}
where $\mathcal{F}_\rmn{\emph{{LWA}}}(\rp \to \infty)$ gives the fraction of galaxies passing our \emph{LWA} selection in the absence of pair interactions probed by our sample selection.  These galaxies are mostly interlopers, associated in projection with truly interacting pairs: a galaxy with \emph{LWA} interacting with one physically close companion could also have additional, non-interacting, companions at any $\rp$.  These companions would be (falsely) counted as interacting in Eqn.~\ref{interlopereq}.  Some galaxies may also display \emph{LWA} features due to intrinsic properties of the galaxies or due to interactions with companions possessing masses lower than our sample selection.  Both these cases are accounted for by subtracting $\mathcal{F}_\rmn{\emph{{LWA}}}(\rp \to \infty)$ in Eqn.~\ref{finteq}. 
%We judge the latter is frequently the case from our inspection of the images in Section \ref{probpairs}.

The factor $P_\rmn{int obs}$ converts the fraction of galaxies in pairs displaying \emph{LWA} into the fraction of truly interacting galaxies.  It is the average probability of a true physical interaction resulting in an observable \emph{LWA} signature in our dataset.  In principle, $P_\rmn{int obs}$ could be a function of $\rp$, although we expect it to vary slowly.  Remember that our working definition of an `interaction', from Section \ref{intro}, is that a galaxy has experienced a significant tidal force, compared to its gravitational binding force, averaged over the previous dynamical time.  For a given dataset, the level of tidal force that is deemed `significant' is that which results in observable features in the most favourable circumstances.  Variations of $P_\rmn{int obs}$ from a constant with $\rp$ are therefore only expected from secondary effects.  However, this assumption would greatly benefit from being tested with simulations, which would potentially result in a refined functional form for $P_\rmn{int obs}$.  For the time being, we assume a constant $P_\rmn{int obs}$.
Although the \emph{LWA} class is a rather indirect indicator of tidal tails, $P_\rmn{int obs}$ accounts for the difference between the number of objects actually counted and the number missed.  Future refinements to Pintobs could included a more careful consideration of the conditions under which interactions produce tidal tails that would be classified as loose spirals. 

%Normally, in this situation, correction factors would be estimated to account for the difference between the number of objects actually counted and the number missed.

\begin{figure}
  \includegraphics[width=60mm,angle=270]{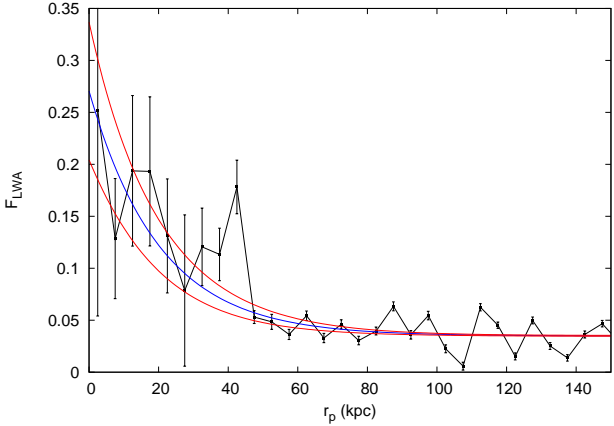}
  \caption{Pairs containing at least one member with $f(\rmn{\emph{LWA}}\,|\, p(\rmn{\emph{FNS}}) > 0.1) > 0.6$ divided by the total number
of pairs in each $\rp$ bin. Results are given for $\Delta V < 500$ \kms (black line), together with a fit using equation \ref{fiteq} (blue line) and the corresponding $1\sigma$ confidence region (red lines).}\label{interloperfig}
\end{figure}

Figure \ref{interloperfig} plots $\mathcal{F}_\rmn{\emph{{LWA}}}(\rp)$ and a fit to the trend using equation \ref{fiteq}. The best fit gives parameter values $a_{\mathcal{F}_\rmn{\emph{{LWA}}}}=0.236$, $b_{\mathcal{F}_\rmn{\emph{{LWA}}}}=20.145$ \hkpc and $c_{\mathcal{F}_\rmn{\emph{{LWA}}}}=0.035$.
At large $\rp$ the curve levels off to a constant value, $\mathcal{F}_\rmn{\emph{{LWA}}}(\rp \to \infty) = c_{\mathcal{F}_\rmn{\emph{{LWA}}}}$.  Therefore, 
\begin{equation}
  \mathcal{F}_\rmn{int}(\rp) = \frac{a_{\mathcal{F}_\rmn{\emph{{LWA}}}} \exp(-\rp/b_{\mathcal{F}_\rmn{\emph{{LWA}}}})}{P_\rmn{int obs}} \,.
\end{equation}

We may place strong constraints on  $P_\rmn{int obs}$ by noticing that $\mathcal{F}_\rmn{int}(\rp)$ must lie on the interval $[0, 1]$.  Therefore, a constant $P_\rmn{int obs}$ must lie on the interval $[a_{\mathcal{F}_\rmn{\emph{{LWA}}}}, 1]$, and hence
\begin{equation}
  \mathcal{F}_\rmn{int}(\rp) = [a_{\mathcal{F}_\rmn{\emph{{LWA}}}}, 1] \cdot \exp(-\rp/b_{\mathcal{F}_\rmn{\emph{{LWA}}}}) \,,
\end{equation}
where the square brackets denote an interval.
This provides the means to calculate strong limits on the true frequency of interactions from our observations of galaxies with \emph{LWA} features.

We can determine an estimate of the average number of interacting companions for galaxies in our sample, ${\Nc}_\rmn{int}(\rp)$, by applying an $\rp$-dependent weight, $w_\rmn{int}$, in the methodology of Section \ref{nccalc}.  This weight is simply the fraction of companions at $\rp$ that are truly interacting galaxies, so $w_\rmn{int} =  \mathcal{F}_\rmn{int}(\rp)$.  For our sample, using the parameters of the fit to $\mathcal{F}_\rmn{\emph{{LWA}}}(\rp)$ in Fig.~\ref{interloperfig}, we derive the interval,
\begin{equation} \label{winteq}
  w_\rmn{int} = [0.236, 1.0] \cdot \exp(-\rp / 20.145\, h^{-1}_{70}\, \rmn{kpc}) \,.
\end{equation}

Usually when one calculates the close companion frequency, one must apply an $\rp$ limit.  Using $w_\rmn{int}$, however, weights companions by the likelihood that pairs with that projected separation are truly interacting.  One can therefore integrate to large $\rp$ and determine a total frequency of interacting companions.  To illustrate, without the interaction weighting (i.e. $w_\rmn{int} = 1$ in Eqn.~\ref{weights}), considering pairs with small projected separations gives $\Nc(\rp < 30\, h^{-1}_{70}\, \rmn{kpc}) = 0.028 \pm 0.002$.  Including the $w_\rmn{int}$ given by Eqn.~\ref{winteq} results in 
${\Nc}_\rmn{int}(\rp < 30\, h^{-1}_{70}\, \rmn{kpc}) = [0.0033 \pm 0.0010, 0.014 \pm 0.004]$, where the interval brackets the allowed range of probabilities, $P_\rmn{int obs}$, that an interaction results in an observable \emph{LWA} feature.
Comparing $\Nc$ and the upper limit for ${\Nc}_\rmn{int}$ implies that at least $49 \pm 14$ per cent of pairs with $\rp < 30$ \hkpc, $\Delta V < 500$ \kms and $M_* > 10^{9.5} M_{\sun}$, are non-interacting interlopers. This agrees with the results of \citet{patt2000}, which found using visual classifications that $\sim 50$ per cent of galaxies were interlopers for a similar range of $\rp$, $\Delta V$ and $M_*$. 

If we select all pairs up to our limiting projected separation, we find that $\Nc(\rp < 300\, h^{-1}_{70}\, \rmn{kpc}) = 0.62 \pm 0.004$, i.e. most galaxies have a companion within this distance.  On the other hand, the average number of physically interacting companions per galaxy is ${\Nc}_\rmn{int}(\rp < 300\, h^{-1}_{70}\, \rmn{kpc}) = [0.0048 \pm 0.0014, 0.021 \pm 0.006]$.  This implies that $> 96$ per cent of pairs with $\rp < 300$ \hkpc, $\Delta V < 500$ \kms and $M_* > 10^{9.5} M_{\sun}$ are not actually interacting.  Our method enables one to estimate the fraction of interacting galaxies without requiring an arbitrary cut-off in projected separation, and without any need for further contamination corrections.

Note that while we have developed a more sophisticated treatment of $\rp$ trends here, we are still using a simple cut in $\Delta V$.  In principle, the method outlined here could possibly be extended to determine an interaction weight, $w_\rmn{int}$, with dependence on both $\rp$ and $\Delta V$.  There is also the potential of further constraining the interaction frequency by combining multiple indicators of interaction, rather than \emph{LWA} alone.  This approach could perhaps also enable the identification of pairs at different stages in their interaction, or with specific orbital characteristics.  We leave all of these challenges for future work.  A paper using the methods outlined above, to explore the dependence of the interaction frequency on stellar mass and environment, is in preparation.

\section{Summary and Discussion}\label{disc}

In this paper we have examined a variety of morphological signatures of interaction between galaxy pairs, and demonstrated how the trends in these observable features, as a function of projected separation, can provide a refined estimate of the frequency of pair interactions in the galaxy population.  We consider an `interacting' galaxy to be one which has experienced a significant tidal force, compared to its gravitational binding force, averaged over the previous dynamical time (see Section \ref{intro}).  The tidal force deemed `significant' depends upon the properties of a given observational dataset.

We began by presenting our sample and the methods we employ, and particularly, in Section \ref{morphprob}, discussing the information provided by Galaxy Zoo 2 and its interpretation in terms of the probability that a given galaxy is observed to possess a particular set of morphological features.  We also presented a method to correct for `projection bias', an effect whereby the signal of certain morphological features may depend on the apparent separation of galaxy pairs, even in the absence of any possible physical associations between pair members.

In Section \ref{oddclass} we considered questions from GZ2 designed to identify \emph{Odd} features, including answers, such as \emph{Merger}, which were intended to identify signs of interaction. We found that these classifications suffer from a strong projection bias.  For example, galaxies with small projected separation, but very large velocity offsets, tend to have a spuriously large \emph{Merger} signal.  Furthermore, the way in which the question for these \emph{Odd} features was arranged, allowing only one of the available options to be selected, results in cross-talk between the \emph{Odd} categories, in terms of both true signal and projection bias, which complicates their interpretation.
We find that this projection bias is also present for the GZ1 \emph{Merger} classification, and shows a behaviour very similar to its GZ2 equivalent. 
Previous studies which used the GZ1 \emph{Merger} class to identify merger candidates \citep{darg2010a,darg2010b} will have suffered to some degree from this issue, but the effect is probably relatively small due to their use of vote fraction thresholds and the fact that the low-$\Delta V$ galaxy pairs have a larger mean \emph{Merger} vote fraction at most projected separations relative to the control sample pairs.
In future iterations of Galaxy Zoo, and other visual classification efforts, it would be preferable to keep questions regarding different types of features distinct, or allow multiple answers to be selected for a single question when the relevant features are not mutually exclusive.

Nevertheless, the \emph{Odd} classifications do provide useful information on the reality of galaxy interactions, particularly once a correction for the projection bias is applied by reference to a control sample of pairs with large velocity offsets.  As discussed by \citet{darg2010a} with relation to GZ1, the GZ2 \emph{Merger} class (and also \emph{Odd=Yes}) primarily selects interacting galaxies at small projected separations.  It therefore mainly probes close passes and the later stages of mergers.  We further find that the \emph{Irregular} and \emph{Disturbed} classes can identify interacting galaxies with very small projected separations, which may be either at an advanced stage of merging or aligned along the line-of-sight.

We have searched all the GZ2 classifications for trends with projected separation, and find significant signals with respect to the presence and form of spiral arms.  The observability of spiral arms (\emph{Spiral=Yes}), and particularly the dominant \emph{2 Arms} class, is enhanced for close pairs on a scale of $\rp \la 70$ \hkpc.  \citet{darg2010a} find that the spiral-to-elliptical ratio for galaxies classified as mergers in GZ1 is approximately twice the global ratio, and in \citet{darg2010b} conclude that this is due to the longer time-scale over which spiral mergers are detectable compared to elliptical mergers. Our results show that this ratio can also be at least partly explained by the enhancement and formation of spiral arms in interacting galaxies.  More unusual spiral arm features also present a trend with $\rp$.  The occurrence of \emph{One Arm} spirals dramatically increases for small separations (on a scale of $\rp \la 20$ \hkpc), while \emph{Loose Winding Arms} show the strongest increase (operating on an intermediate scale of $\rp \la 30$ \hkpc). 
 
There are two principal ways in which spiral-like features can be created through tidal interactions. Tidal perturbations can instigate or amplify instabilities in gas disks, leading to the formation or enhancement of star formation in spiral arms similar to those seen in isolated galaxies (e.g., as seen by \citealt{xu2010}). Tidal forces can also strip stars and gas out of the galaxies, forming tidal tails, counter tails and bridges, which may or may not harbour star-formation (e.g., \citealt{mull2011}).  We appear to detect both signatures: an enhancement of `normal' spiral arm features, occurring at large projected separations, with the signatures of stronger tidal interactions becoming increasingly prevalent at smaller separations.  The tidal nature of the \emph{Loose Winding Arms} features is confirmed by examination of typical images, such as those in Table \ref{lwaimages}.  It is clear that many of the galaxies which have significant probability of \emph{Loose Winding Arms}, especially those with higher stellar mass, are red, early-type galaxies. This indicates that the loose spiral features that are observed in these galaxies, and probably also those same features in star-forming, late-type galaxies, are the result of tidal stripping.  In galaxies with sufficient cold gas, there will almost certainly be star formation in these tidal spiral features, and indeed in Table \ref{lwaimages} we see that several of the galaxies possess very blue loose spiral arms.  Simulations (e.g., \citealt{toom1972,howa1993,barn2011}) indicate that tidal features, such as those apparently identified by the GZ2 \emph{Loose Winding Arms} class, are indicative of a stage between close passes, primarily between 1st and 2nd pass, when pairs can still attain relatively large separations. The \emph{Loose Winding Arms} features are thus probing the early stage of mergers and pair interactions.

The onset and appearance of tidal arms are known to depend on the geometry of the encounter.  Numerical studies demonstrate that in-plane, prograde encounters produce the most symmetrical two-sided disturbances, while polar encounters give the most one-sided disturbances, and retrograde encounters are the last to make tidal tails (e.g., \citealt{thom1989,howa1993,barn2009,barn2011}). Retrograde encounters also produce the greatest increase in star formation efficiency \citep{cox2008}.  Considering this, the galaxies which are selected as having \emph{2 arms} and \emph{Loose Winding Arms} in GZ2 are likely the result of prograde, in-plane encounters, while galaxies identified as displaying \emph{1 arm} are likely the result of polar or retrograde encounters.  Our observed separation scales for these different features are consistent with this interpretation (see Table~\ref{fitparam}).

When comparing our results to other studies which look at the onset of tidally induced changes in interacting galaxies, we find that answers to Galaxy Zoo questions regarding spiral arms detect changes at separations similar to studies of tidally induced star formation, as discussed in Section \ref{intro}. The \emph{Loose Winding Arm} class begins to detect interacting galaxies around $\rp \la 120$ \hkpc, which is similar to the separation scale associated with induced star formation \citep{niko2004,li2008,patt2011}. The star formation detected at these large separations is relatively weak, while a strong increase is observed for $\rp \la 40$ \hkpc \citep{li2008,elli2008,roba2009,patt2011}. This corresponds to scale for which we observe an enhancement of the \emph{Merger} class.  Quantitative morphological measurements also typically present signals on these scales \citep{hern2005,depr2007,elli2010}. Logically, kinematic disturbance must precede morphological disturbance and so it might seem that star formation, if it is triggered by kinematical perturbations, should be an earlier indication of interaction than morphology (e.g., \citealt{byrd1992}).  However, this paper shows that some morphological signatures are as sensitive as enhanced star-formation, and more unambiguously related to interaction.

In Section \ref{bars} we found that the likelihood of a bar being observed decreases sharply for pairs with projected separations $\rp \la 20$ \hkpc.
Bars are thought to be created through periodic orbital resonance \citep{bour2002} and are known to initiate radial gas inflows, which in the end act to destroy or weaken the bar structure \citep{pfen1990} (for a recent review see \citet{sell2010}, or more comprehensively \citet{sell1993}).
 Gas inflows, perhaps together with the enhancement of bar features, caused by tidal perturbations in the early stages of major interactions may similarly act to rapidly destroy any pre-existing or transient bars (e.g., \citealt{bere2003,dima2007}).  Our results suggest that this is indeed the case, with the appearance of bars being strongly suppressed in close pairs, in agreement with other recent studies by \citet{mend2011} and \citet{lee2012}.  Eventually, the violent reorganization of stellar orbits in the later stages of many major interactions (i.e. mergers) must act to erase any orbital resonances which created the bar.

In Section \ref{probpairs} we focus on using the presence of \emph{Loose Winding Arms} to identify probable interacting galaxies.  These criteria are then used in Section \ref{interlopers} to constrain the frequency of galaxy pair interactions, without requiring an arbitrary cut-off in projected separation or any further corrections for contamination of our close pair sample.  We find that the fraction of galaxies with $M_* > 10^{9.5} M_{\sun}$ and $0.01 < z < 0.05$ that are in truly interacting pairs with $\Delta V < 500$ \kms is in the range $0.5 \pm 0.1$--$2.1 \pm 0.6$ per cent. The limits correspond to assuming the maximum and minimum permitted probability, $P_\rmn{int obs}$, that interacting galaxies produce observable \emph{Loose Winding Arms} features, respectively.  We expect simple extensions of our technique to lead to significantly tighter confidence intervals in future work.

It is difficult to precisely compare our estimate of the interacting galaxy fraction to other studies, due to the range of different methods employed.  Although the close pair fraction is mostly constant with luminosity \citep{patt2008}, the limiting mass-ratio and projected separation, varying definitions for selecting pairs, and many other subtleties, make comparisons complicated.   Given that the estimate in this paper is derived from a relatively simple demonstration of combining close pair and morphological information, we defer such involved comparisons to future work.  Nevertheless, we note that the major interaction fractions quoted by other recent studies: e.g., $1.1 \pm 0.5$,  $2.1 \pm 0.1$,  $1.6 \pm 0.1$,  $1.3 \pm 0.1$ by \citet{bell2006,patt2008,domi2009,xu2012}, respectively, are neatly bracketed by our estimate of 0.4 -- 2.7 per cent.  This lends support for our use of GZ2 \emph{Loose Winding Arms} as indicators of pair interactions, and encourages confidence in the method described in Section \ref{interlopers}, and the various assumptions we have made.

\section*{Acknowledgments}

This publication has been made possible by the participation of more than 200000 volunteers in the Galaxy Zoo project. Their contributions are individually acknowledged at http://authors.galaxyzoo.org .

Much of the work presented was performed while KRVC was hosted at the University of Nottingham, on a visit funded by a research travel grant from the Government of Catalunya, Spain (ref. 2010 BE-00268).
KRVC would also like to thank Ben Hoyle for introducing him to Galaxy Zoo and getting him started on this project.

SPB gratefully acknowledges receipt of an STFC Advanced Fellowship.

KLM acknowledges funding from The Leverhulme Trust as a 2010 Early Career Fellow.

Support for the work of KS was provided by NASA through Einstein Postdoctoral Fellowship grant number PF9-00069 issued by the Chandra X-ray Observatory Center, which is operated by the Smithsonian Astrophysical Observatory for and on behalf of NASA under contract NAS8-03060. 

Finally we thank the referee, Curt Struck, for his helpful comments and suggestions in preparing this work for publication.

\label{lastpage}

\end{document}